\documentstyle[11pt,aaspp4]{article}

\def\beq{\begin{equation}}
\def\eeq{\end{equation}}
\def\beqar{\begin{eqnarray}}
\def\eeqar{\end{eqnarray}}

\def\pcite#1{(\cite{#1})} 
\def\pref#1{(\ref{#1})}   

\def\bline{\rule[1.2mm]{3em}{0.1mm}}

\def\la{\mathrel{\mathpalette\fun <}}
\def\ga{\mathrel{\mathpalette\fun >}}
\def\fun#1#2{\lower3.6pt\vbox{\baselineskip0pt\lineskip.9pt
  \ialign{$\mathsurround=0pt#1\hfil##\hfil$\crcr#2\crcr\sim\crcr}}}

\def\li#1{\hbox{${}^{#1}$Li}}
\def\be#1{\hbox{${}^{#1}$Be}}
\def\bor1#1{\hbox{${}^{1#1}$B}}

\def\feh{\hbox{$[{\rm Fe/H}]$}}
\def\oh{\hbox{$[{\rm O/H}]$}}

\def\avg#1{\langle #1 \rangle}
\def\msol{\hbox{$M_{\odot}$}}

\def\lesc{\Lambda_{\rm esc}}
\def\grams{\, {\rm g} \, {\rm cm}^{-2}}

\def\mgas{M_{\rm gas}}
\def\esn{\epsilon_{\rm SN}}
\def\eism{\epsilon_{\rm ISM}}

\def\fia{\hbox{$f_{\rm Ia}$}}
\def\yldz{{\cal Y}_{\rm O}}
\def\imf{\xi}

\def\dbefe{\hbox{$\omega_{\rm BeFe}$}}
\def\dbeo{\hbox{$\omega_{\rm BeO}$}}
\def\dbfe{\hbox{$\omega_{\rm BFe}$}}
\def\dbo{\hbox{$\omega_{\rm BO}$}}
\def\dofe{\hbox{$\omega_{\rm O/Fe}$}}
\def\dij#1#2{\hbox{$\omega_{#1#2}$}}
\def\etal{{\rm et al.}~}

\def\nupro{$\nu$-process}

\begin{document}
\rightline{  }
\vskip -1in
\rightline{UMN-TH-1719/98}
\rightline{astro-ph/9809277}
\rightline{September 1998}

\title{THE REVIVAL OF GALACTIC COSMIC RAY NUCLEOSYNTHESIS?}

\author{Brian D. Fields}
\affil{Department of Astronomy , University of Illinois \\
Urbana, IL 61801, USA \\
}

\author{Keith A. Olive}
\affil{School of Physics and Astronomy, University of Minnesota \\ 
Minneapolis, MN 55455,  USA \\
}
            
\begin{abstract} 
Because of the roughly linear correlation between Be/H and Fe/H in low
metallicity halo stars, it has been argued that a ``primary'' component in
the nucleosynthesis of Be must be present in addition to the ``secondary''
component from standard Galactic cosmic ray nucleosynthesis.   In this
paper we critically re-evaluate the evidence for
the primary versus
secondary character of Li, Be, and B evolution, 
analyzing both in the
observations and in Galactic chemical evolution models.
While it appears that [Be/H] versus [Fe/H] has a logarithmic
slope near 1, it is rather the Be-O trend that 
directly arises from the physics of spallation production.
Using new abundances for oxygen in halo stars based on UV OH lines, we
find that in Pop II stars for which O has been measured, 
the Be-O slope has a large uncertainty due to systematic effects.
Namely, the Be-O  logarithmic slope lies in the range 
$1.3 - 1.8$, rendering it
difficult to distinguish from the data between the secondary slope of 2
and the primary slope of 1.  The possible difference between the Be-Fe
and Be-O slopes is a consequence of the variation in O/Fe versus Fe:  
recent data suggests that the best-fit O/Fe-Fe slope for Pop II
is in the range
-0.5 to -0.2, rather than zero (i.e., Fe $\propto$ O) as is often assumed.
In addition to this phenomenological analysis of Be and B evolution,
we have also examined the predicted LiBeB, O, and Fe trends
in Galactic chemical evolution models which include outflow.
Based on our results, it is possible that 
 a good fit to the LiBeB
evolution requires only traditional the Galactic cosmic ray spallation,
and the (primary) 
neutrino-process contribution to \bor11.  We thus suggest that
these  two processes might be sufficient to explain \li6, Be, and B
evolution in the Galaxy, without the need for an additional primary
source of Be and B.
However, the uncertainties in the data at this time
prevent one from reaching a definitive conclusion.   
Fortunately, several observational tests of this  ``neoclassical'' 
scenario are available; we note in 
particular the importance of further observations to 
secure the O/Fe Pop II trend, as well as
accurate measurements of B/Be, \li6/Be, and \bor11/\bor10 in halo stars.
\end{abstract}


\section{Introduction}

Galactic cosmic rays (GCR) have long been known to be 
a significant source of lithium, beryllium, and boron 
nucleosynthesis
(Reeves, Fowler, \& Hoyle \cite{rfh}; 
Meneguzzi, Audouze, \& Reeves \cite{mar}).
These elements are produced via spallation and fusion
reactions between cosmic ray nuclei and those in the 
interstellar medium (ISM).
Indeed, until the 1990's it was thought that
LiBeB are
{\em predominantly} synthesized in this way
(except for \li7 and perhaps \bor11).
Estimates of the accumulation of
LiBeB over the age of galaxy, using
the observed, present-day GCR flux and ISM abundances
(e.g., Walker, Mathews, \& Viola \cite{wmv}),
can reproduce 
roughly the right {\em solar system} values for the
absolute abundances and the 
isotopic ratios of \li6, \be9, and \bor10.
Some question remained about \li7 and \bor11,
which seemed to require additional nucleosynthesis sources,
but the agreement was seen as a confirmation of the
basic GCR nucleosynthesis process as the
source of LiBeB.

Our picture of LiBeB production 
was partially confirmed and partially challenged 
in the past decade, when
data became available on LiBeB {\em evolution}
in prior epochs.
Specifically, observations revealed the
pattern of elemental abundances of LiBeB versus Fe
in both disk and halo (Population I and II) stars.
While elemental Li is dominated at low metallicity
by the ``Spite plateau'' arising from the primordial component, 
Be and B show no such plateau, but rather scale with metallicity.
Thus, Be and B have no
significant primordial component as was to be expected from standard
models of big bang nucleosynthesis (Thomas \etal \cite{tsof},
Delbourgo-Salvador \& Vangioni-Flam \cite{dsvf}), and have a  Galactic
production site, in accordance with the basic ``galactogenetic''
hypothesis of Reeves, Fowler, \& Hoyle
\pcite{rfh}.

However, the same stellar data that showed Be and B to have a Galactic origin 
also threw into question the ``standard'' LiBeB nucleosynthesis 
scenario of GCR spallation on ISM nuclei.
The observed Be- and B-Fe relations show logarithmic
slopes (for brevity, ``slopes'')
much closer to 1 than to 2.  
However, in standard GCR nucleosynthesis,
Be and B are ``secondary'' 
since the production rate of Be and B depends on the abundance of
the ``primary'' target nuclei C, N, and O in the ISM.
Thus, in this scenario,  
Be and B have slope 2 versus, say, oxygen.\footnote
{
The GCR production of Li is more complicated, since primary production of
\li6 and \li7 via $\alpha+\alpha \rightarrow {\rm Li}$ occurs in 
addition to CNO spallation (e.g., Steigman \& Walker \cite{sw}).
}
Therefore,
{\em if O and Fe abundances are proportional to one another} in Pop II 
stars, 
the Be and B data are in conflict with 
standard GCR nucleosynthesis.
The data have thus been interpreted as requiring
both primary Be and 
primary B.  In particular, when one normalizes to the solar abundances,
the difference between the primary and
secondary production is largest at low metallicities and hence
early epochs.
Thus the Be and B data in halo (Pop II) stars apparently suggest a need for 
other nucleosynthesis sites in the early Galaxy for these element
isotopes.

Of the primary mechanisms suggested, most involve
accelerated particle interactions, but one---the ``neutrino process''---is 
a stellar mechanism (Woosley \etal \cite{w1}; Olive \etal \cite{opsv},
Woosley \& Weaver \cite{ww}).
The \nupro\ occurs as the onion-skin layers of a supernova
are traversed by the intense neutrino flux.
Inelastic neutrino collisions with nuclei in the 
carbon and helium shells can lead to the production of 
\bor11 and \li7.  This mechanism is a ``primary'' one,
in that a ``seed'' abundance is not needed in the supernova's
progenitor for the \nupro\ to occur.
Since most current stellar abundance data measures only 
the elemental abundance of B, the \nupro\ alone could
explain the B-Fe trend, though adjustments are needed
to fit the solar \bor11/\bor10 
(Olive, Prantzos, Scully, \& Vangioni-Flam \cite{opsv}, Vangioni-Flam,
Cass\'{e}, Fields, \& Olive \cite{vcfo}).

Even if the \nupro\ is responsible for the
Pop II trends in B, the $\nu$-process makes no Be.  Thus, the 
observed trend of Be versus Fe, with slope near 1, 
apparently demands an additional primary source of Be
which dominated in the early Galaxy.
One proposed mechanism invokes a flux of 
accelerated particles that are localized to star forming regions
(Cass\'{e}, Lehoucq, \& Vangioni-Flam \cite{clv};
Ramaty, Kozlovsky, \& Lingenfelter \cite{rkl}).
These particles would be
energetic (i.e., above the thresholds for LiBeB spallation production),
but non-relativistic, and dominated by C and O nuclei.
This mechanism is in part motived by $\gamma$-ray observations of 
${}^{12}{\rm C}^{*}$ and ${}^{16}{\rm O}^{*}$ lines in Orion,
which are only explained as the result of such a population of
energetic particles.
Another mechanism posits the direct acceleration of 
supernova ejecta (Duncan, Lambert, \& Lemke \cite{dll};
Ramaty, Kozlovsky, Lingenfelter, \& Reeves \cite{rklr}).
In this scenario, the
C and O component is assumed to dominate the composition,
but the energetic particles are now at relativistic energies.

Thus we see that the apparently primary nature of
the B-Fe and Be-Fe trends has driven the need for new LiBeB production
in the early Galaxy.
Given the far-reaching implications of such new production,
it is worthwhile to scrutinize the empirical basis 
of the primary behavior.  
Here we wish to re-examine critically 
whether the B-Fe and Be-Fe slopes necessarily imply 
primary origins for B and Be.  In fact, a better
and  more direct indicator of the 
nucleosynthetic origin of Be and B is oxygen.  
This is well known, but it has been thought that O/Fe is constant in Pop
II, so that Fe can act as a surrogate for O. However, we will discuss
recent data which suggests that O/Fe may not be constant.  Using recent
Pop II oxygen data for halo stars, we find that the Be-O and B-O slopes at
present  suffer from systematic
effects, the result of which is
that the slopes may lie in the range 1.3 -- 1.8.
Interestingly, this range of uncertainty lies
between the values of 1 and 2 arising for
primary and secondary origins, respectively.
Thus it is unclear on the basis of these data alone
whether primary or secondary production occurred in the
early Galaxy.

In the face of this uncertainty, we reconsider in this paper
the standard GCR nucleosynthesis scenario, in which
\li6 and \li7 are effectively primary in Pop II
(due to $\alpha+\alpha$ reactions;
Montmerle \cite{mont}; Steigman \& Walker \cite{sw}),
and \be9 and \bor10 are secondary.
Of course, since the observed meteoritic
\bor11/\bor10 isotopic ratio cannot
be fit by standard (relativistic) GCR processes, one
must add something to this scenario to fit even the solar
abundances.  
We will adopt the point of view that the
$\nu$-process  does occur, and must contribute to \bor11 (and probably
\li7 as well). Indeed, we know supernovae are intense neutrino sources,
and so spallative production must occur at some level.
Of course, the solar isotopic data on \li7 and \bor11
demand that some additional process, other than conventional GCR spallation,
produces at least these isotopes.  
If we take the $\nu$-process yields seriously, then
this contributes a primary component to Pop II \bor11 
(and thus elemental B), which in turn demands that Be and B slopes
differ, and thus that B/Be rises toward low metallicities
(Olive \etal \cite{opsv}, Fields, Olive \& Schramm \cite{fos95}).
Given the $\nu$-process,
there would not be a strong requirement for other (primary) sources of B.
Thus, the only evidence driving the need for additional primary sources
of LiBeB is the Be data.

In what follows we will review the BeB data with respect to both Fe and O.
We find that the data is particular vulnerable to systematic effects
which depend on the stellar atmospheric parameters such as surface
temperature and gravity. Due to these uncertainties, it is unfortunately
not possible with the present data to determine unambiguously the primary
versus secondary nature of Be.  We then ask, if Be indeed has an origin in
standard GCRs throughout the history of the Galaxy, what are the
implications for LiBeB and cosmic ray  evolution if this is true?  
We find that simple chemical evolution models can obtain good fits
to Be and B vs O and Fe.  Acceptable BeB vs OFe evolution can arise
not only in simple closed box scenarios, but also in open box models
which allow for Galactic outflows (Scully et al.\ \cite{sean}).
We find that although the effect of Galactic winds is to
flatten the Be evolution, the effect is not significant enough to change
the slope by one unit.
Finally, we point out further empirical tests which can
determine whether a new and separate cosmic ray component is required to
operate in the  early Galaxy.

\section{Data}

\subsection{LiBeB}

In order to model and test existing models of the synthesis and evolution
of the LiBeB isotopes, it is essential to have reliable abundance data. To
derive abundances from the observed line strengths, one must adopt a model
stellar atmosphere models. In practice, reported abundances use different
assumptions  in the model atmospheres for all stars; 
e.g., whether one imposes local thermodynamic equilibrium (LTE)
or not (NLTE).  In addition, somewhat different parameters 
are adopted for the same star, e.g., surface gravity, $T_{\rm eff}$, 
or \feh.
To meaningfully derive and compare trends for different elements
across metallicities, it is thus crucial to systematize
the stellar atmosphere models that underly the 
abundances.  In what follows, where multiple measurements exist for
a given star, we will combine them systematically at a common set of
stellar parameters to determine the abundances of Be/H, B/H and O/H as
was done in Vangioni-Flam \etal (\cite{vroc}, hereafter VROC). 
Unfortunately, the choice for these parameters is not unique and their
choice can have a significant effect on the evolutionary trends for Be
and B as we show below.

Having a dominant primordial component at low metallicities, \li7 differs
from the other element isotopes we are considering. For our
purposes here, Galactic LiBeB production is constrained by \li7 only in
that models must respect the Spite plateau in metallicity. In most cases,
this is not a problem.  There is of course an observed rise in the \li7
abundance at [Fe/H] $>$ -1, but we will not address the issue of 
late \li7 production here. We will assume a Spite plateau value of 
Li/H $= 1.6 \times 10^{-10}$ (Molaro, Primas, \&  Bonifacio \cite{mpb},
Bonifacio \& Molaro \cite{bm}).

We will also not go into detail concerning \li6 here (the question
of \li6 will be treated in Vangioni-Flam, Cass{\'e}, Olive \& Fields
\cite{vcof}).  We do note here however, that the $\alpha -
\alpha$ fusion process which produces \li6 in addition to spallative
processes (see e.g. Steigman \& Walker \cite{sw}), is effectively a
primary process.  
If Be was in fact a secondary element as in the standard
model of GCR nucleosynthesis, the \li6/\be9 ratio
would change with metallicity as appears to be the case when one
compares this ratio in halo stars (see e.g. recent observations of \li6
by Smith, Lambert, \& Nissen \cite{sln2}, Cayrel \etal \cite{cay}) to the
solar value.

The logarithmic abundances of Be and B scale with those of heavy elements.
There is no evidence as yet of a primordial component, which from the
standard model of big bang nucleosynthesis is expected to be unobservably
small (Thomas \etal \cite{tsof}, Delbourgo-Salvador \& Vangioni-Flam
\cite{dsvf}) given current observational capabilities. The Be and B data
can be fit versus \feh\, using logarithmic abundances so that,
\beq
\label{eq:slope_def}
[{\rm Be}] = \dbefe \, \feh \ + \ const
\eeq
We will focus on the logarithmic slope $\dbefe$;
the same procedure, applied to boron, gives $\dbfe$.
[Be] is defined as $\log$(Be/H) + 12.

To determine the slopes $\dbefe$ and $\dbfe$, one needs to choose a
consistent set of stellar parameters. One choice advocated by 
Molaro, Primas, \&  Bonifacio \pcite{mpb} to obtain reliable \li7
abundances, is to use surface temperatures as given by
Fuhrmann, Axer, \& Gehren
\pcite{fag} based on Balmer lines. This approach was used in VROC for
obtaining the Be-Fe and B-Fe correlation. The specific BeB data used for
these fits was discussed in VROC and we refer the reader there for more
details. Here we have also adjusted the surface gravity and iron
abundances to match those given by Axer, Fuhrmann, \& Gehren
\pcite{afg}. These results for the Be-Fe and B-Fe slopes for [Fe/H]
$<$ -1,  appear in Table
\ref{tab:data} and are very similar to those obtained in VROC. 
However, this choice of stellar parameters based on Balmer line
observations is not unique.  An alternative is to use data based on
the Infra-Red Flux Method (IRFM) of Alonso, Arribas, \& Martinez-Roger
\pcite{aam1,aam2}.  This was used by Bonifacio \& Molaro \pcite{bm} for
\li7 as well and gave very consistent results (with Balmer line method)
for the \li7 abundance in the Spite plateau.  The IRFM of Alonso, Arribas, \& Martinez-Roger
\pcite{aam2} was used by
Israelian, Garcia Lopez, \& Rebolo \pcite{iglr} to obtain oxygen
abundances in halo stars, and we have also adopted these parameters to
adjust the Be and B abundances so that we again are utilizing a
consistent parameter set. The results for the Be and B slopes based on the
IRFM are also shown in Table \ref{tab:data}.

As one can
see, the Be and B show a clear departure from  the quadratic (slope = 2)
dependence on [Fe/H] commonly expected in standard GCR nucleosynthesis
for both the Balmer line and IRFM parameter choices. In fact these results are
entirely consistent with one another.  
In the case of B/H, there are two points
at low metallicity which stand out due to an NLTE correction that has been
applied.  If we use only the LTE abundance data for B/H (though we have
no reason to suspect a problem with the NLTE corrections that have been
applied) the slope becomes $\dbfe = 0.93 \pm 0.13$ and is not
qualitatively different than the NLTE case. We will discuss the Be-O and
B-O trends below.

As in the case of \li6, the isotopic ratio \bor11/\bor10 in Pop II stars
is determined via isotopic splitting, in this case at
$2090 \, {\rm \AA}$.  Also similar to \li6, this is a difficult
observation. Rebull et al.\ \pcite{rdjtf} made a first attempt at
measuring this in HD 76932, which has $\feh = -1$, and $\oh =
-0.62\pm0.23$ (Israelian, Garcia Lopez, \& Rebolo \cite{iglr}) and found
a possible blended line. If the 2090 \AA\ feature is unblended and
entirely due to B, then
$\bor11/\bor10 \sim 4-10$.  If an (unknown) blend
is present, then the ratio could be as low as $\bor11/\bor10 \sim 1$.
Further work is needed to address the issue of blending.

\begin{table}[htb]
\caption{Observed Pop II logarithmic slopes for Be and B versus Fe and O}
\begin{tabular}{cccccc}
\hline\hline
 metal tracer & method & metallicity range & Be slope & B slope & B/Be
slope
\\
\hline
Fe/H & Balmer & $-3 \le \feh \le -1$ & $1.21 \pm 0.12$ & $0.65 \pm 0.11$
&     $-0.18 \pm 0.15$ \\
O/H & Balmer  & $-2.5 \le \oh \le -0.5$ & $1.76 \pm 0.28$ & $1.84 \pm
0.58$ & $-0.81 \pm 0.44$ \\
Fe/H & IRFM & $-3 \le \feh \le -1$ & $1.30 \pm 0.13$ & $0.77 \pm 0.13$
&     $0.01 \pm 0.14$ \\
O/H & IRFM  & $-2.5 \le \oh \le -0.5$ & $1.38 \pm 0.19$ & $1.35 \pm 0.30$
& $0.00 \pm 0.17$ \\
\hline\hline
\end{tabular}
\label{tab:data}
\end{table}

As we will discuss below, the solar abundances of \li6BeB
are sufficient to fix all of the parameters of LiBeB production,
and thus these abundances are key inputs.
For the solar Li, Be, O, and Fe abundances, we will adopt the
values of Anders \& Grevesse \pcite{ag}.
For the elemental B, we note the uncertainty in the solar value,
and will note the impact of using the meteoritic value 
${\rm B/H} = (6.06 \pm 0.79) \times 10^{-10}$ of 
Zhai \& Shaw \pcite{zs}
(recommended by Grevesse, Noels, \& Sauval \cite{gns}),
or the photospheric value 
${\rm B/H} = (4.0 \pm 2.8) \times 10^{-10}$. For the boron isotopes,
we use the Chaussidon \& Robert \pcite{cr} value
$\bor11/\bor10 = 4.05 \pm 0.16$.

\subsection{Oxygen Data and Trend Versus Iron}  
\label{sect:o/fe}

Oxygen provides a more direct measure of the LiBeB nucleosynthetic origin
than does Fe, 
since O (along with C and N) is the nucleus that is fragmented to make LiBeB.
In practice, however, O is not as common a metallicity standard
as Fe since O is harder to measure. 
Indeed, there has been debate over which O lines
are the best indicators of the true oxygen abundance,
since use of different lines have led to systematically different
O/H abundances.
Abundances determinations have used several indicators:
(1) the allowed \ion{O}{1} triplet at 7774 \AA;
(2) the forbidden [\ion{O}{1}] line at 6300 \AA; and
(3) molecular OH lines at 3085 \AA.
The third method was suggested by Bessell, Hughes \& Cottrell \pcite{bhc}
and recently high resolution UV spectra were obtained for many of the
stars with Be and B observations (Nissen \etal \cite{ngeg}, Israelian,
Garc\'{\i}a-L\'{o}pez, \& Rebolo
\cite{iglr}).
These new data suggest a significant nonzero [O/Fe] trend (versus \feh)
in Pop II stars. 
We should point out that there are systematic uncertainties in the [O/Fe]
ratio and the magnitude of these are difficult to estimate. A comparison
of the different methods to measure O/H and the results of different
measurements was made by Israelian,
Garc\'{\i}a-L\'{o}pez, \& Rebolo
\pcite{iglr}. Hereafter, we will adopt the O abundances
derived from the OH lines, and examine the
implications of the results.

The O/Fe trends versus Fe in Pop I is well-known to show
a decrease in O/Fe from
$\feh=-1$ to $\oh=0$.  However, for 
Pop II, the data has been less clear.  
O/Fe has often been assumed to be
constant at low metallicity, and the data was uncertain enough 
to allow this possibility.  
However, the data of
Israelian, Garc\'{\i}a-L\'{o}pez, \& Rebolo \pcite{iglr}
strongly suggests that O/Fe is not constant, but rather
increases towards low metallicity.
Specifically, Israelian et al.\ find 
that for Pop II metallicities 
$-3 \le \feh \le - 1$ 
(or $-1.8 \la \oh \la -0.5$),
\beq
\label{eq:o_fe}
[{\rm O}/{\rm Fe}] = \dofe \feh + const
\eeq
with $\dofe = - 0.31 \pm 0.11$, 
based on 17 stars.
Thus we have 
\beqar
\oh & = & [{\rm O}/{\rm Fe}] + \feh \\
\label{eq:o_h}  
& = & (1 + \dofe) \ \feh
\eeqar
The new evidence that $\dofe \ne 0$
would mean, via eq.\ \pref{eq:o_h}, 
that the common assumption that
${\rm O} \propto {\rm Fe}$ would be
incorrect for Pop II.

 The [O/Fe] data relevant to the Be and B data considered here is taken
from the OH data of Nissen \etal \pcite{ngeg} and Israelian,
Garc\'{\i}a-L\'{o}pez, \& Rebolo
\pcite{iglr} and is plotted in Figure \ref{fig:ofe_fia}.
A fit to [O/Fe] versus [Fe/H] yields a very steep slope of 
\beq
\dofe = -0.46 \pm 0.15 \qquad {\rm (Balmer)}
\label{ofeb}
\eeq
based on data from 20 stars.
This slope differs from zero at the $3\sigma$ level.
On the other hand, using the IRFM for the same stars, we have
\beq
\dofe = -0.22 \pm 0.15 \qquad {\rm (IRFM)}
\label{ofeirfm}
\eeq
which differs from zero only in a statistically mild way.
These bracket and are consistent statistically with the slope of $-0.31
\pm 0.11$ quoted in  Israelian, Garc\'{\i}a-L\'{o}pez, \& Rebolo
\pcite{iglr}.
The slopes in Eqs. (\ref{ofeb}) and (\ref{ofeirfm}) include the data of
Nissen \etal \pcite{ngeg} and in the Balmer line 
case we have applied a systematic
shift in the stellar parameters used
in order to that they match those chosen for the Be and B data,
where we used a uniform set of
stellar data.  In this
way when we compare Be/H versus O/H, the data for all abundances are
based on the the same stellar parameters.  The predominant effect in
the difference for the oxygen abundances is the choice of surface
temperature. The data for both the Blamer line method and IRFM are shown
in Figure 1.

If $\dofe$ differs from 0, then the usual assumption
that in Pop II, ${\rm O} \propto {\rm Fe}$ not longer holds.
In turn, the usual argument no longer necessarily holds  
that the Be- and B-Fe slopes near $\sim 1$ implies a primary origin
for Be and B.
That is, although the slopes against
Fe/H do seem consistent, and lean to the primary interpretation, such a
conclusion would be fallacious unless we had conclusive evidence that the
O/Fe-Fe slope were in fact 0.  
Until this is established, we will
need a better understanding of the 
element abundances of the light elements
as well as O before we can answer the question regarding the
primary/secondary nature of Be and B.
In what follows, we outline how a nonzero O/Fe-Fe slope
changes the usual analysis of Be and B origin.

\section{Be and B Trends Versus Oxygen and Implications for Be and B Origin}
\label{sect:beb-o}

Once the validity of ${\rm O} \propto {\rm Fe}$ for Pop II stars
has been called into question, the Be and B versus Fe slope
no longer remains a direct indicator of the nucleosynthetic origin
of LiBeB.  Thus it becomes important to determine the Be and B
slopes versus O.  Two ways to determine the slopes
$\dbeo$ and $\dbo$ are as follows.  
(1) Examine the Be and B trends versus O in stars where these are measured
as discussed in the previous section.
The obvious advantage of this approach is that it directly addresses
the issue at hand.  The disadvantage, however, is that presently the 
needed data exist for fewer Pop II stars than have
Be or B and Fe measured.  Thus the slopes determined by this method
have larger errors due to the smaller sample size.   (2) To reduce
these errors, one can take a different approach, which
supplements the measured Be and B versus Fe slopes with
the measured O-Fe trend to arrive at Be and B versus O.
We will take both approaches and compare their results.

The simplest approach conceptually is to
determine directly the Be and B versus O trends for Pop II.
As one can see in Table \ref{tab:data}, 
the slopes of Be and B versus O/H are far from certain.
For both, the range for the slopes is roughly 1.3 -- 1.8, placing them
between what is expected if they are primary or secondary elements.
These results intriguingly call into question the
presupposition that the slopes should show a value near
1, which would indicate a primary origin.
If anything,
the slopes versus oxygen may even favor a secondary origin.
However, this analysis alone cannot definitively 
discriminate between the two hypotheses of BeB origin,
as the data are still quite 
uncertain (and not nearly as numerous as the BeB set
versus iron).
Indeed, it is the systematic difference in the results when the two
sets of stellar parameters are chosen that implies that at this time
based on this data, it is not possible to determine the primary/secondary
nature of beryllium and boron.    Still, one can at the least say 
that the BeB data versus oxygen certainly offers no
overwhelming support for a primary origin.

In the face of these uncertainties, 
an alternative to using Be, B versus O directly is to use
the better determined BeB trends versus Fe, along with 
the O-Fe data,
to infer Be and B versus O for Pop II.  While this approach
is less direct, it has the obvious advantage of the increased size of
both the O and the BeB
data sets. 
Another advantage to this method is that one can
directly compare the Be and B slopes versus Fe, and the
differences are independent of the O-Fe trend 
(as long as there is {\em some} trend).
Consider an element ${\cal A}= \be9$, \bor10, or \bor11.
If ${\cal A}$ has some fixed log slope $\dij{\cal A}{, {\rm O}}$ versus O,
\beq
[{\cal A}] = \dij{\cal A}{, {\rm O}} \, \oh + const
\eeq
and if we use eq.\ \pref{eq:o_h} to relate $\oh$ to $\feh$
in Pop II stars,
we have
\beq
[{\cal A}] = \dij{\cal A}{, {\rm O}} \, (1 + \dofe) \feh
\eeq
and so
\beq
\label{eq:Fe_vs_O_slope}
\dij{\cal A}{, {\rm Fe}} 
  = \dij{\cal A}{, {\rm O}} \, (1 + \dofe) 
\eeq
The upshot of eq.\ \pref{eq:Fe_vs_O_slope} is that
the iron slope $\dij{\cal A}{, {\rm Fe}}$ does not
directly reveal the nucleosynthesis origin of ${\cal A}$,
but depends on both the true indicator $\dij{\cal A}{, {\rm O}}$
and the $\dofe$ slope.

For example, consider the case in which
${\cal A}$ is primary versus O and the O/Fe-Fe
slope is not zero, but takes the
Israelian, Garc\'{\i}a-L\'{o}pez, \& Rebolo
\pcite{iglr} value $\dofe = -.31 \pm 0.11$, which lies between
the values of Eqs. (\ref{ofeb}) and (\ref{ofeirfm}).
Then $\dij{\cal A}{, {\rm O}} \equiv 1$ by assumption, and 
by eq.\ \pref{eq:Fe_vs_O_slope} the iron slope is
$\dij{\cal A}{, {\rm Fe}}  = 1 + \dofe = 0.69 \pm 0.11$. 
This ``primary''
slope is entirely consistent with the observed
$\dbfe$ vales for both the
Balmer line and the IRFM data sets, 
and thus suggests that B truly has a ``primary'' origin
versus oxygen (contrary to the weak indications by the B-O slope
of a secondary origin).

On the other hand, 
in the case where ${\cal A}$ is secondary versus O, then
$\dij{\cal A}{, {\rm O}} \equiv 2$, which gives an iron slope
$\dij{\cal A}{, {\rm Fe}}  = 2(1 + \dofe) = 1.38 \pm 0.22$.
This predicted ``secondary'' slope is considerably less than 2.
Furthermore, this predicted slope {\em agrees} with
the observed
$\dbefe$ within the total error budget.
That is, for both the Balmer line and the IRFM 
a sets, an inferred secondary Be-Fe slope 
is within 
$1\sigma$ of the observed Be-Fe value,
while and inferred 
primary slope differs
from the observed $\dbefe$ by $> 3 \sigma$.
On this basis, one could argue that of the eight slopes listed for Be and
B in Table 1, only one shows a departure from the expected behavior. If we
had assumed that Be was secondary while B was primary (with respect to
O/H), then apart from the IRFM Be-O slope the others are consistent
(the B-O slope also shows a departure from expectations, but the
uncertainty is very large in this case).

Also note that if ${\cal A}$ is secondary in O and another
element ${\cal B}$
is primary in O, then the difference in the slopes
gives the slope of the ${\cal B/A}$ ratio, and is
inferred to be
\beq
\dij{\cal B/A,}{\rm Fe}
   = \dij{\cal B}{\rm Fe} - \dij{\cal A}{\rm Fe} = - 1 - \dofe
   = - 0.69 \pm 0.11
\eeq
thus, primary and secondary oxygen slopes (which differ by 1 unit)
lead to an iron slope difference $< 1$. 
In other words, if B is primary in Pop II, and Be secondary,
then one expects B/Be to vary with Fe, but
not as strongly as the na\"{\i}ve slope difference of 1.
Interestingly, the Table \ref{tab:data} data
is ambiguous regarding whether the 
Be and B slopes differ, depending
on the metal tracer used.  
For both data sets, the Be and B slopes versus Fe
are different at the $ 3.5\sigma$ level for the 
Balmer set, and the $2.9\sigma$ level for the IRMF set.
On the other hand, the (more uncertain) Be and B slopes versus O are
entirely consistent with each other for both sets,
despite the large difference in the O slope values between the
Balmer and IRFM sets.  Thus, the Be and B data versus Fe
seem to require a variation in B/Be, while the
more uncertain data versus O allows for some variation
but does not at all demand it.
These results are to be compared with the B/Be slope
as obtained directly from individual stars.
In fact, such direct determinations are 
few--only 6 Pop II data points are available, thus making 
direct B/Be slope determinations very uncertain.
Indeed, as seen in Table \ref{tab:data}, 
the direct data on B/Be are ambiguous on
whether the slope is nonzero; the Balmer line set allows
this, while the IRFM set disfavors a strong evolution.
As we will see below using specific models, the difference
in B/Be between these two sets is enough to allow
for standard GCR nucleosynthesis in the Balmer case,
but to disfavor it in the IRFM case.  This again points to the
significance of the current uncertainties, and the
need for more and better data.

To summarize, we see that the (better measured) 
slopes versus Fe
strongly hint at different origins
for Be and B, while the slopes versus O
are ambiguous due to their uncertainties.
In particular, the BeB-Fe slopes in conjunction
with a nonzero O/Fe-Fe slope suggests that 
B is primary in
O, while Be appears to be secondary.  
Emboldened by this suggestion with the data, we now
turn to specific models to examine this scenario of
LiBeB origin.

\section{Galactic Chemical Evolution and Cosmic Ray Nucleosynthesis}

\subsection{Chemical Evolution Models}

We model Galactic chemical evolution following
the standard formalism described well elsewhere (e.g.,
Tinsley \cite{tins}).  Thus we will briefly 
note the basic equations in order to 
establish conventions.

For the one-zone models we will consider, 
the total mass varies as
\beq
\label{eq:mtot}
\frac{d}{dt} M_{\rm tot} = - \vartheta
\eeq
where we have allowed for an ``open box'' with a gas outflow
rate $\vartheta$.  We will also consider the simple
``closed box'' case in which $\vartheta = 0$.
The gas mass changes via
\beq
\label{eq:mgas}
\frac{d}{dt} \mgas = - \psi + E - \vartheta
\eeq
where $\psi$ is the star formation rate.
The total mass ejection rate from dying stars is
$E = \int \, dm \, m_{\rm ej} \, \imf(m) \, \psi[t-\tau(m)]$
depends on the star formation rate as well as the initial mass function
$\imf$ and the lifetime $\tau$ of stars of mass $m$.
A nuclear species $\ell$ has a gas mass fraction
$X_\ell = M_{{\rm gas},\ell}/\mgas$ which evolves according to
\beq
\label{eq:massfrac}
\mgas \, \frac{d}{dt} X_\ell = E_\ell - X_\ell E
   - (X_\ell - X_\ell^{\vartheta})\vartheta 
   + Q_\ell  
\eeq
Here
$E_\ell = \int \, dm \, m_{{\rm ej},\ell} \, \imf(m) \, \psi[t-\tau(m)]$
where $m_{{\rm ej},\ell}$ is the mass ejected from stars in the form of
$\ell$. For the case of LiBeB, $Q_\ell$ is the mass production rate from
cosmic rays (derived below, eq.\ \pref{eq:Q_CR}).
Note that because LiBeB are very fragile, they are destroyed
by $(p,\alpha)$ reactions in all of the convective zones of 
stars; thus, the bulk of stellar ejecta are LiBeB free, and
so we take $E_\ell = 0$ for these elements.  The exception to this
occurs for \bor11 and \li7 production by the neutrino process,
for which we have $E_\ell = E_{{\rm SN},\ell} = E_{\nu,\ell}$.

Following, e.g., Scully et al.\ \pcite{sean},
we consider gas outflows having two components.
One type is a bulk outflow in which ISM material is heated
by multiple supernovae,
and driven out of the Galaxy by an evaporative wind.
In the simplest model (Hartwick \cite{hart}),
the outflow strength is taken to be proportional to the supernova
rate and thus the star formation rate:
$\vartheta = \eism \psi$.
The bulk outflow
composition is usually assumed to be the same as that
of the ISM, i.e.,
$X_\ell^{\vartheta} = X_\ell$.
The other type of outflow we consider
is motivated by a picture in which supernova blast waves
preferentially expel supernova products from the
Galaxy (Vader \cite{vader}).  In this ``enriched'' outflow scenario,
the flow strength is taken to be a fraction
of the total supernova ejecta only:
$\vartheta = \esn E_{\rm SN}$.
Furthermore, the composition is exactly that of 
the supernova ejecta:
$X_\ell^{\vartheta} = X_\ell^{\rm ej,SN} = E_{{\rm SN},\ell}/E_{\rm SN}$.
While the ejection efficiency could in principle be different
for type Ia and type II supernovae, for simplicity we will
assume the factor is same for both.
Note that since $\nu$-process yields are of SN origin,
these are included in the enriched winds.

Since we are interested in iron evolution in general, and
O/Fe in particular,
it is essential to include Type Ia supernovae, which
contribute significantly to the Fe abundance.
To do this we follow the prescription of
Matteucci \& Greggio \cite{mg}; 
these authors take Type Ia origin 
to be accretion onto a white dwarf from
a companion.
Thus, the Type Ia rate is computed via integrals over
the total mass $M$ as well as the mass fraction
$\mu$ of the companion:
\beq
\label{eq:typeIa}
R_{\rm Ia} = \fia \ \int dM \ \imf(M) \ 
   \int \, d\mu \ g(\mu) \ \psi\left[t - \tau(m_2)\right]
\eeq
where $m_2 = \mu M$ is the mass of the secondary,
and we follow Matteucci \& Greggio
in taking the distribution $g(\mu) \propto \mu^2$.
For both prescriptions,
the nucleosynthesis yields are then modified
by adding a term $E_i^{\rm Ia} = M_i^{\rm Ia} R_{\rm Ia}$,
where the yields $M_i^{\rm Ia}$ are tabulated by
Thielemann, Nomoto, \& Yokoi \pcite{tny}.

\subsection{LiBeB Nucleosynthesis Models}

The formalism for cosmic ray nucleosynthesis is discussed
in detail in Fields, Olive, \& Schramm \pcite{fos94}.
To summarize, the  production rate $\lambda_{ij}^{\ell}$
of LiBeB species $\ell$ 
per target atom of species $i$
via the process $i + {j} \rightarrow \ell + \cdots$ is
\beq
\lambda_{ij}^{\ell} = \int \ dE \ \sigma_{ij}^{\ell} \ \phi_j \ S_{\ell} \\
\eeq
where $\phi_j$ is the flux of energetic nuclei of species $j$,
$\sigma_{ij}^{\ell}$ is the cross section for
$i + j \rightarrow \ell$.  The dimensionless factor
$S_{\ell}$ measures the probability of stopping 
the LiBeB daughter against escape from the Galaxy:
$S_{\ell} \simeq 1$ for the processes 
${p,\alpha}+{\rm CNO} \rightarrow {\rm LiBeB}$, but
$S_{\ell}$ is noticeably less than unity
for the ``reverse'' kinematics (fast CNO on interstellar H and He).

We will implement standard GCR nucleosynthesis as follows.
We assume that the cosmic ray flux (in species $j$) is separable
in energy and time,
with 
$\phi_j(E,t) = \varphi_j(E) \Phi_j(t)$.
As described in Fields, Olive, \& Schramm \pcite{fos94},
the energy spectrum is propagated (for each species $j$
from a source spectrum 
$q(E) \propto (E+m_{\rm p})^{-2.7}$, consistent
with current data.  Since we are modeling standard GCR nucleosynthesis,
the only low-energy ($<100$ MeV) cosmic rays are those which arise from a
smooth extrapolation of the high energy flux.  For simplicity, we use a
constant $\lesc = 11
\grams$. The spectra are normalized via $\int dE \, \varphi_j(E) = 1$,
and the energy dependence is essentially the same for all species of
interest over the energy ranges of interest.  Finally,
we define a time independent, spectrum-averaged cross section
$\avg{\sigma_{ij}^{\ell}} = \int dE \, \varphi(E) \sigma_{ij}^{\ell}$;
using this, the production rate becomes 
$\lambda_{ij}^{\ell}(t) = \Phi_j(t) \avg{\sigma_{ij}^{\ell}}$.

We further separate the time dependent part of the flux
to isolate the composition and the flux strength.
Namely, we write $\Phi_j(t) = y^{\rm CR}_j(t) \, \Phi(t)$,
where $y^{\rm CR}_j(t) = \Phi_j/\Phi \, (t)$ is the cosmic ray composition
and $\Phi(t)$ (without subscript) is the total integrated
proton flux.  For standard GCR nucleosynthesis, we 
assume that the cosmic rays arise as supernova shocks accelerate
the ambient ISM.  Thus, we assume that
(1) the cosmic ray composition is the same as 
that of the ISM, 
i.e., $y^{\rm CR}_j(t) = y^{\rm ISM}_j(t) = X_j/A_j X_{\rm H} \, (t)$.
We also assume that (2) the cosmic ray flux is proportional to the
Type II supernova rate:  $\Phi \propto R_{\rm II}$.  
To a good approximation,
$R_{\rm II} \propto \psi$, so that we have 
$\Phi = \Phi_0 \psi(t)/\psi_0$, where subscript 0 denotes
the present Galactic average value.

To summarize, we have 
\beqar
\lambda_{ij}^{\ell}(t) & = & 
   \Phi(t) \ \sum_{ij} y^{\rm CR}_j(t) \ \avg{\sigma_{ij}^{\ell}} \\
  & \propto & 
       \left\{ \begin{array}{ll}
                      \psi(t) & \mbox{Pop II \li6,\li7} \\
                      X_{\rm CNO}(t) \ \psi(t) & \mbox{Pop II Be,\bor10}
              \end{array} \right.
\eeqar
The heavy element target abundances $X_{\rm CNO}$ lead to 
the secondary nature of Be and \bor10, while
the lack of this factor due to $\alpha+\alpha$ gives a primary
nature to GCR \li6 and \li7 (as discussed in more detail below).

To include GCR nucleosynthesis in chemical evolution, we must relate the
local production rate per atom, $\lambda_{ij}^{\ell}$, to the Galactic
mass production rate, $Q_\ell$. The production rate of the number of
nuclei of species
$\ell$ per unit volume is
\beq
j_\ell \ \equiv \ \left( \frac{d}{dt} n_\ell \right)_{\rm CR} 
   \ = \ \sum_{ij} \lambda_{ij}^{\ell} n_i 
\eeq
or in terms of mass density,
\beqar
\nonumber
q_\ell & = & m_\ell \ j_\ell \\
\nonumber
   & = & \sum_{ij} \ \frac{m_\ell}{m_i} \ \rho_i \  \\
   & = & \rho_{\rm gas} \  \Phi(t) \ 
      \sum_{ij} \ \frac{m_\ell}{m_i} \ X_i \ y^{\rm CR}_j \ 
      \avg{\sigma_{ij}^{\ell}}  
\eeqar
The sum has a strongly time varying factor in the target
abundances $X_i$. If we put $z_i \equiv X_i/X_{\rm O}$,
we can define an effective cross section 
$\bar{\sigma}_\ell = \sum_{ij} \, ({m_\ell}/{m_i}) \, z_i \, y^{\rm CR}_j
\, 
      \avg{\sigma_{ij}^{\ell}}$ which varies slowly with time 
due to the variations in C:N:O.
We can then write
\beq
q_\ell = \rho_{\rm gas} \  \Phi \ X_{\rm O} \ \bar{\sigma}_\ell
\eeq  
Integrating over a homogeneous volume $V$, 
we thus have a LiBeB mass production rate
\beq
Q_\ell = \left( \frac{d}{dt} M_\ell \right)_{\rm CR} \ 
   = \mgas \  \Phi \ X_{\rm O} \ \bar{\sigma}_\ell
\label{eq:Q_CR}
\eeq

\section{Results}

\subsection{Analytic Results}

As usual, one cannot solve the full chemical evolution equations 
analytically,
but one can obtain illustrative analytic results by adopting the instantaneous
recycling approximation (IRA).
In fact, within the IRA, one can solve for LiBeB evolution 
(Pagel \cite{pagel}, \cite{pagel94}; 
Ryan et al.\ \cite{rnbd}) over the full
range of metallicity, but the solution is complicated
and can obscure basic simple result.
Instead, we will focus on the limiting cases of metallicity,
where ``metallicity'' $Z = X_{\rm O}$ here is understood to be 
the mass fraction of O and not Fe.
At low metallicities, the destruction term is much
smaller than production term, $X_\ell \psi \ll \Upsilon_\ell \psi Z \mgas$,
where $\Upsilon_\ell = \Phi_0 \bar{\sigma}_\ell Z_0 \tau_{\rm SF,0}$
is the cosmic ray production (without astration) over
the star formation timescale $\tau_{\rm SF,0} = M_{\rm gas,0}/\psi_{0}$
with present ISM metallicity $Z_0$.
When $X_\ell$ small, i.e., at low metallicity,
we have
$\dot{X}_\ell \simeq \Upsilon_\ell \psi Z \mgas$
and 
\beq
\label{eq:slope-loZ}
\frac {d X_\ell}{d Z} \propto Z \mgas
\eeq
which holds generally, with no use of the IRA.
The $\mgas$ factor in eq.\ \pref{eq:slope-loZ} leads to 
a deviation from the simple quadratic relation,
as emphasized by Yoshii, Mathews, \& Kajino \pcite{ymk}.  

To proceed, we use the closed box IRA result
$\mgas \propto e^{-Z/\yldz}$, where $\yldz$ is the yield of $Z$
(see, e.g., Tinsley \cite{tins}).
With this, we can solve:
\beqar
X_\ell & \propto & 1 - (1+Z/\yldz) e^{-Z/\yldz} \\
  & = & \frac{1}{2} \left( \frac{Z}{\yldz} \right)^2
      - \frac{1}{3} \left( \frac{Z}{\yldz} \right)^3 + \cdots
\eeqar
We see that LiBeB is indeed quadratic at lowest metallicities, but
a deviation from the quadratic quickly grows 
as metallicity increases.\footnote
{
In fact, the flattening of the slope at higher metallicities
occurs regardless of whether the isotope $\ell$ is primary or secondary.
This supports our decision to focus on the slope at low
metallicity, where the deviation due to the $\mgas$ factor
is minimized.
}
To first order in the deviation, the log slope is
$\dij{\ell}{\rm O} = 2 - 2Z/3\yldz$ for secondary production
(and $\dij{\ell}{\rm O} = 1 - Z/2\yldz$ for primary production).
Thus, the BeB slope versus O
is slightly smaller than the na\"{\i}vely expected value.
The size of the departure depends on the value of the yield,
which itself depends on the IMF as well as the inclusion
of any winds.  For a closed box as we have assumed for illustration,
$\yldz \sim 0.01  \sim  X_{\rm O,\odot}$,
so the log slope is reduced to
1.93 at mass fraction 
$Z = 0.1 \yldz$.  For an open box, the yield is
effectively smaller, and so the slope is {\em flatter}
at a fixed metallicity $Z$.  This is in part the motivation for
considering outflow in this context.

The other limit of the LiBeB evolution occurs at higher metallicities, where
a steady state is reached:
$d X_\ell/dt = 0$. 
The steady state
abundance is easily obtained as
\beqar
\label{eq:L-hiZ}
X_\ell^{\rm ss} & \propto & \left( \frac{\psi}{E} \right) \ Z \mgas \\
   & \stackrel{\rm IRA}{=} & Z \ e^{-Z/\yldz}
\eeqar
where the first expression holds generally (with
$E$ the gas ejection function of eq.\ \ref{eq:mgas}),
and the latter expression holding in the IRA
(in which $\psi/E$ is constant).
Both expressions are
valid where $Z > \yldz$; in this regime
$X_\ell^{\rm ss}$ {\em decreases}, since the scarcity of target nuclei in the
dwindling gas reservoir more than offsets the 
rising gas metallicity.  The lesson of the analytical results,
therefore, is that a BeB production
has a log slope that only approaches the na\"{\i}ve values
a the lowest metallicities, and then flattens and finally
turns over, with LiBeB actually declining at late times.
Moreover, this flattening is generic and occurs
{\em regardless} of whether the production is primary or secondary.

While the analytic, IRA solutions provide valuable
intuition, the
assumptions therein are not obeyed for several 
elements which are central
to our study,
and which arise from low-mass, long-lived stars.
These elements include
not only Fe, but also the C and N that comprise
spallation targets.
We thus turn to more realistic results
from numerical models.

\subsection{Numerical Results}

We have implemented
the full chemical evolution 
equations (eqs.\ \ref{eq:mtot}--\ref{eq:massfrac}),
without the use of the IRA,
and solve them numerically.
The basic model ingredients are described in 
Fields \& Olive \pcite{fo};
most notably, the stellar nucleosynthesis yields
are metallicity-dependent, and
use the tabulations of van den Hoek \& Groenewegen \pcite{vdhg}
for $1-8 \msol$ stars, and
Woosley \& Weaver \pcite{ww} for $11-40 \msol$ stars
(including their neutrino process yields).

\subsubsection{The O/Fe Ratio}
\label{sect:ofe}

As discussed in \S\S \ref{sect:o/fe}-\ref{sect:beb-o},
any chemical evolution model constructed to fit the BeB slopes versus Fe
and O must also reproduce the nonzero O/Fe slope
in Pop II.  
In modeling O/Fe, the yields of both elements of
course play central roles, but O production
is simpler--Type II supernovae are the only important source--and
the O yields of different groups are in good agreement 
(Woosley \& Weaver \cite{ww}; 
Thielemann et al.\ \cite{tnh}).
In contrast, the production of Fe is more complicated,
since Type Ia and Type II supernovae both make
important contributions. Furthermore, there are
significant model dependences in how to include the
Type Ia yields in chemical evolution, 
and there are significant uncertainties
in the Type II Fe yield.  Consequently,
we will first explore how the predicted O/Fe slopes
depend on the uncertainties
in Fe production.

To illustrate the basic effect of Type Ia SNe on
iron evolution, we have run models in which we
varied the SNIa ``amplitude'' $\fia$ given in Eq. (\ref{eq:typeIa}).
Here we follow the Matteucci \& Greggio \pcite{mg}
SNIa prescription of eq.\ \pref{eq:typeIa}, 
for a closed box model, with an IMF $\imf \propto m^{-2.35}$.
The resulting plot of O/Fe appears in Figure \ref{fig:ofe_fia}.
Note that the curves for different $\fia$ are
all the same until some minimum
metallicity, $\feh \simeq -2$, where they begin to diverge.
The slopes below $\feh \simeq -2$ are entirely
due to Type II SNe, as these stars produce all of the O and Fe at
early epochs until the 
most massive companions ($\le 8 \msol$)
of SNIa progenitors
can evolve off the main sequence.
The $\fia = 0$ curve corresponds to Type II yields only;
we see that here that [O/Fe] is flat at
high metallicity, contrary to the data, and does not go to 
zero (i.e., solar) at $\feh = 0$.
We see that the inclusion
of the Ia's leads to a smooth transition
from the Pop I flatness of the SN II--only curve
to an unbroken descent from the low-metallicity slope.

For our purposes, Figure \ref{fig:ofe_fia}
implies that the O/Fe slope below $\feh \simeq -2$ is
entirely a consequence of the SNII yields.
This means that modification of the SNIa scheme
can only influence the $\feh \ga -2$ regime, 
while the lowest metallicity region
 can only reflect SNII and their inclusion
in the chemical evolution model.
Specifically, the problem is that the
predicted O/Fe slope is too shallow (i.e., less negative)
than the data show.  This slope reflects the change
in the mean SN II yields as a larger range of Type II
progenitor masses is sampled over time.
Thus, the steepness of the observed slope suggests
that the O/Fe ratio was higher in Type II ejecta
at high masses.  

One possible solution to this discrepancy
lies in the yields themselves;
as noted above, the O yields are the more secure, 
so we focus on the Type II yields of Fe.
To illustrate the effect of varying the Type II yields,
we have explored one alternative that has some physical
justification.  Namely, we have run models with a 
mass of ejected iron which is constant for all the progenitor masses; 
this is motivated by the view that
the physics of the supernova core is roughly independent
of the progenitor mass.  Specifically, we adopt the
observed value from SN 1987A, $m_{\rm ej,Fe} = 0.07 \msol$.
Indeed, we find that this gives a somewhat better fit to the O/Fe 
data, with slope $\dofe = -0.20$, as compared to $\dofe = -0.18$ for
the Woosley \& Weaver \pcite{ww} yields of Fe (though the two
are not very different, since the full yields indeed do not vary
strongly with progenitor mass).  

Another model feature which affects the calculated O/Fe slope
is the IMF.  Different IMF slopes $x$ 
(with IMF $\imf \propto m^{-x}$) change
the contribution of higher-mass to lower-mass
SNII progenitors, as well as the SNIa/SNII ratio.
In particular, the smaller $x$ favors more high-mass SNII,
as well as more SNII per SNIa; both of these effects
go in the direction of increasing O/Fe, particularly at low
metallicity.
In fact, we find that a smaller $x$ does
increase O/Fe over all metallicities (to give a much better fit
to the solar point), but the
Pop II slope is in fact slightly shallower:
we get $\dofe = -0.20$ for $x=2.35$
versus -0.24 for $x=2.7$.
In addition to the IMF slope and the SN yields 
other model features, such
as a sequential or 
bimodal IMF, or outflow,
cant significantly change the predicted O/Fe behavior.
Indeed, we find that a sequential model, as in
Scully et al.\ \pcite{sean} can give an improved
O/Fe fit; the results for a bimodal model lie between
those for the sequential and time-independent IMFs.  
Consequently, we are left with a situation in which
reasonable choices of supernova yields and IMF slopes
can qualitatively reproduce the basic observed O/Fe-Fe trend--namely
a steady decline from a high value at low metallicity.
However, the available yields and IMFs do not reproduce
the steep decline of O/Fe in quantitative detail.

In summary, 
we find that although chemical evolution models can
naturally get a varying O/Fe ratio in Pop II, with
the yields we use, we generally can not reproduce the O/Fe-Fe
slopes we see in the Balmer data set.  
Of course, the theoretical BeB slopes versus O are unaffected
by O/Fe difficulties (so long as the O yields are indeed accurate).
However, given that the BeB-O data suffers from
greater uncertainties than the data versus Fe,
it is clearly worthwhile (and conventional) to also
show the evolution versus Fe.
To do this, we must circumvent our inability to 
quantitatively fit O/Fe-Fe in our {\em ab initio} calculation.  
In order to obtain more realistic BeB slopes versus Fe,
we will not compute Fe from SN yields, but instead
use the observed O/Fe-Fe slope in conjunction with the
O evolution as computed from the more reliable yields.
Thus we will put $\feh = \oh/(1 + \dofe)$.
In this way, we still rely on our code to compute the
evolution histories we believe are simple 
(i.e., those of \li6BeB and O), but for the more complicated
case of Fe, we will use the observed trend versus
the (simpler) O abundance rather than our model results.
Thus, while we do not solve the (possible) problem of
Pop II evolution of O/Fe, we can still get a feel for
the problem at hand,
namely how the LiBeB evolution would look in a model
with O/Fe in agreement with the data.

\subsubsection{Be and B Slopes and Nucleosynthetic Origin}

We have included GCR and $\nu$-process nucleosynthesis
in our chemical evolution models, and now examine
Be and B evolution in the GCR nucleosynthesis scenario.  
As noted in the previous section, 
in the case of Fe we will use the computed O evolution,
and the observed O/Fe slope, to derive the Fe evolution.
This means the BeB slopes versus O are genuine outputs of 
our code, while the slopes versus Fe are not ``independent''
outputs, but are related to the O slopes via 
eq.\ \pref{eq:Fe_vs_O_slope}.\footnote
{
However, in practice, the O and Fe metallicity ranges of the 
fits cover epochs that are not identical, which means that
the Fe slopes are not precisely what would follow
from eq.\ \pref{eq:Fe_vs_O_slope}.
}

There is a question of normalization in
the absolute abundances of LiBeB.  While
this should in principle be fixed by the
present cosmic ray total flux 
(and thus present LiBeB production rate),
we acknowledge uncertainties in this quantity
by normalizing the final abundances of GCR-only
isotopes (\li6, \be9, and \bor10) to have
a solar abundance at $\feh = 0$,\footnote
{
Specifically, we find the scaling factor for
each of these isotopes separately, and use
the average of the three.
}
we also adjust the GCR yields of \li7 and \bor11
by the same factor.  This scaling amounts to a 
determination of the (poorly determined)
Galactic average of the present cosmic ray
flux $\Phi_0$ (e.g., Vangioni-Flam et al.\ \cite{vcfo}).
Finally, we adjust the 
$\nu$-process yields
by scaling so that \bor11/\bor10 at
$\feh = 0$ is equal to the observed
ratio, $4.05 \pm 0.16$ (Chaussidon \& Robert \cite{cr}).  
This adjustment reflects model uncertainties in
the $\nu$-process yields,
(Olive et al.\ \cite{opsv}; Vangioni-Flam et al.\ \cite{vcfo},
\cite{vroc}), constraining the temperature of the $\nu$ flux.
In all cases, \li7 has
an additional primordial component of 
$\li7/{\rm H} = 1.6 \times 10^{-10}$ 
(Molaro, Primas, \& Bonifacio \cite{mpb}; 
Bonifacio \& Molaro \cite{bm}).

Notice that these normalizations exhaust the freedom
to adjust the LiBeB evolution.  Given the ``neoclassical''
evolution scheme we have adopted--standard GCR plus
$\nu$-process nucleosynthesis--the basic trends
are essentially fixed, and the only model-dependence
lies in the uncertainties due to the chemical evolution
model itself.  Thus we will examine the effect of 
different chemical evolution scenarios (closed box versus
outflow) and input parameters (IMF).
	
For each chemical evolution model, we compute Be and B slopes
numerically, in the same manner
as the data fits of Table \ref{tab:data}.  Namely,
for the O and Fe slopes, we use the set of
model points having $-2.5 \le \oh \le -0.5$
and $-3 \le \feh \le -1$, respectively.
For these sets of points, we calculate
BeB versus O and Fe slopes by a least-squares fit.
We find in practice that the Be and B log-log curves in these
regimes are generally well-fit by a line, so the
slopes are a good description of the data.
Note that the LiBeB scaling, described in the
previous paragraph, does not affect the Be slopes.
However, scaling {\it does} affect the B slopes 
when we include both
the (primary) $\nu$-process and the (secondary)
GCR nucleosynthesis, since the latter dominates
at high metallicities, and the scaling affects
the epoch at which the transition occurs.

\begin{table}
\caption{Slopes for numerical models}

\begin{tabular}{ccccccc}
\hline\hline
Model & $\esn$ & $\dofe$ & $\dbeo$ & $\dbo$ & $\dbefe$ & $\dbfe$ \\ 
\hline
Closed box & 0.00 & $\equiv$ -0.31 & 1.87 & 1.31 & 1.32 & 0.88 \\
Outflow & 0.00 & $\equiv$ -0.31 & 1.94 & 1.26 & 1.34 & 0.86 \\
Outflow & 0.50 & $\equiv$ -0.31 & 1.87 & 1.31 & 1.33 & 0.87 \\
Outflow & 0.60 & $\equiv$ -0.31 & 1.85 & 1.34 & 1.31 & 0.89 \\
Outflow & 0.70 & $\equiv$ -0.31 & 1.79 & 1.40 & 1.32 & 0.92 \\
\hline\hline
\end{tabular} 
\label{tab:runs}
\end{table}

As a benchmark case we consider a 
closed box model.  As we will see below, the quality of
the fit to the depends on the adopted IMF slope $x$,
where we put $\imf \propto m^{-x}$.
We find that for a closed box, the best fit
has an IMF $\imf \propto m^{-2.65}$, which
is a good approximation to high-mass end
of the local disk IMF.
The slopes appear in 
Table \ref{tab:runs}, and
the evolution curves
appear in Figures \ref{fig:BeB-O_cl_b},
\ref{fig:LiBeB-Fe_cl_b},
and \ref{fig:BBe-Fe_b}a
along with the Balmer data.
{}From the table, we see that the Be-O slope is 1.98,
very close
to (but less than) two, as expected for a purely secondary
product.  The B-O slope is near (but larger than) one,
indicating that in the range of the fit, the primary
component of $\nu$-process \bor11 dominates,
but the secondary component has some contribution as well.
As the figures show, the models provide a good fit to the
data for the both the abundances and the ratios.\footnote
{
Aside from the Li behavior at high metallicity, where the
model underproduces \li7.  The $\nu$-process
yields in our scheme are unable to fit the solar \li7;
this need not be a problem as other sources of Pop II \li7 have been
proposed (e.g., AGB stars).
}
Consequently, this example from a full chemical evolution model
supports the conclusion of our phenomenological analysis 
(\S \ref{sect:beb-o}):  it is possible that \li6BeB evolution
can be explained solely by a combination of
standard GCR nucleosynthesis and the $\nu$-process.

Having established the benchmark case, we turn to open box models.  
We will explore the effect adding an enriched wind to
models with a bulk wind.  Here we use a shallower IMF,
$\imf \propto m^{-2.35}$,
which has the needed effect of increasing the stellar yields
to compensate for the metals lost in the wind.
The results are not strongly sensitive to the
strength of the bulk wind, with the best fits coming
from values around $\eism = 0.33$ (with $\eism$ much lower, the
bulk wind has no effect; much higher, and models lose
too many metals and gas).
In contrast, the models are much more sensitive to the
strength of the enriched wind, as seen in Table \ref{tab:runs},
and in Figures \ref{fig:BBe-Fe_b}b, 
\ref{fig:BeB-O_E+B-wind_b}, and \ref{fig:LiBeB-Fe_E+B-wind_b}.
The most important point is that the open box
models do not change the LiBeB evolution dramatically;
the basic trends are very similar.
Still, we do see in Table \ref{tab:runs}
as expected, $\dbeo$ is slightly but systematically
lowered due to the earlier
(in $\oh$) onset of gas depletion.
More importantly, we see that as $\esn$ increases,
the BeB curves turn over at lower metallicity; this
arises due to the lowered effective oxygen 
yield $\yldz^{eff} = (1-\esn)\yldz$, which 
causes the gas mass (and target reservoir)
to drop faster versus O (see eq.\ \ref{eq:L-hiZ}).
The lower turnover improves the fit, since
the curves are forced to go through the solar point,
and a shallowly positive or even negative slope
at this point effectively raises the curve
at the Pop II metallicities.
Finally, we note that the $\esn=0$ case, in comparison
to the closed box case, gives a sense of the effect of 
a change in the IMF.  Here again, the key is the yield,
which is lower in the closed box case (steeper IMF),
leading to the same improvement in fit seen
via lowering the effective yield.

In terms of the overall fit of the outflow model
curves to the Balmer data, 
we see that the $\esn = 0.7$ model does particularly well.
In the Balmer case, the Be slope of about 2 is reduced, and the B slope is
increased which in fact better matches the data. The reduction in the Be
is slope is not however enough to match the IRFM data.  Again, this model
serves as an example of
\li6BeB evolution explained by a GCR + $\nu$-process scenario. 
Furthermore, it is encouraging that one can find a solution 
within a wind model as well as a closed box.  Comparing
the $\esn = 0.7$ and the closed box model, we see that 
the two are quite similar, but the wind model in fact
gives a slower evolution in B/Be, and thus a somewhat better fit
to the few Pop II B/Be points.

The outflow models of Figures
\ref{fig:BBe-Fe_b}b, 
\ref{fig:BeB-O_E+B-wind_b}, and \ref{fig:LiBeB-Fe_E+B-wind_b}.
are plotted along with the IRFM data in
Figures \ref{fig:BeB-O_E+B-wind_i}, 
\ref{fig:LiBeB-Fe_E+B-wind_i}, and \ref{fig:BBe-Fe_b}c.
As anticipated in \S \ref{sect:beb-o},
here the fit to B-O is not very good.
This comes about due to the primary, $\nu$-process component of 
\bor11 which makes too
much boron at early times.  This excess boron
appearly mostly starkly in B/Be.
Indeed, as suggested by the B/Be slopes of Table \ref{tab:data},
the IRFM data shows no evidence for a change in the B/Be slope.
The difference between the fits in Figures
\ref{fig:BBe-Fe_b}b and \ref{fig:BBe-Fe_b}c points
to the seriousness of the systematic errors in the data,
and the inability of the data to discriminate at present
between different schemes for \li6BeB origin and evolution.
By the same token, the difference between
Figures
\ref{fig:BBe-Fe_b}b and \ref{fig:BBe-Fe_b}c
also highlights 
a golden opportunity
for observers:  if the data can be improved, it will be 
become possible to distinguish between the
different evolutionary hypotheses, and to
rule in or out whole classes of light element models.

Interestingly, we find that some chemical evolution models 
which have been invoked to solve the 
G-dwarf problem are not compatible with a good solution
to the O-Fe or BeB-O trends.  
In particular, neither sequential nor bimodal
outflow help to get the large O-Fe demanded by data,
though these models were constructed to give good fits
to the G-dwarf distribution (Scully et al.\ \cite{sean}).
In the case of sequential models, the first burst of star
formation
builds up floor of metals which then act as targets
for the renewed burst of star formation.  The BeB curves thus
have early- and late-time solutions which are unchanged,
but in the transition region BeB grows rapidly, introducing
a ``jump'' in the BeB versus metal curves.  This behavior 
leads to a bad fit, since
it effectively lowers the Pop II trend relative to that of Pop I.
A similar but weaker effect arises in a bimodal model.
We believe that this behavior is characteristic of any model which
has in effect a prompt initial enrichment to solve the G-dwarf problem.

Finally, we note the changed role of Be production 
energetics in our scenario.
The energy production per Be atom
is a powerful diagnostic for models
(Ramaty, Kozlovsky, Lingenfelter, \& Reeves \cite{rklr}), as 
it links the Be production to the needed energy budget.
However, this diagnostic is not 
as useful if the O/Fe slope is not equal to 0.
If Be is secondary versus O/H, its production in this case depends on
the ISM abundances and so the ``energy per atom''
scaling is metallicity dependent: clearly,
the energy that a Pop II supernova invests in cosmic rays
leads to a lower payoff in BeB 
than the same cosmic ray blast from a Pop I supernova, due to the
decrease in number of CNO targets.
The energetics argument requires the near constancy of Be/Fe.
 By changing O/Fe, one could have a Be/H evolution which is
secondary with respect to O/H, but as we have shown, has a slope close to
unity with respect to Fe/H. Even if Be/Fe is roughly constant, it does
not necessary follow that  there is constant energy budget per Be atom.
Indeed, the possibility that O/Fe is not constant has implications beyond
our present study, since this means that at least one of O or Fe is not a
simple surrogate for ``number of supernovae,'' and that one must be
careful in assigning a single ``metallicity'' for a given epoch.

To summarize, our numerical results support conclusions
suggested by the more approximate analytic treatment,
in showing that Be could well have its
nucleosynthetic origin
in the traditional, ``secondary,'' process of
GCR spallation in the interstellar medium.
The situation for B is less clear, largely due to
the paucity of B data, but the B-O slope 
also seems to indicate that
B may have a similar origin in the GCR nucleosynthesis process,
though the differences in the B and Be slopes versus Fe
suggest that B might also have a significant primary
component, presumably the $\nu$-process. 
Fortunately, the scenario we propose here is testable
is several ways, as we now discuss.

\section{Discussion and Conclusions}

We have scrutinized the issue of primary versus secondary Be and B,
examining both the trends in the data itself, and in  models for
LiBeB nucleosynthesis.
Phenomenological, analytical, and numerical
analyses all suggest that \li6, Be, \bor10, and part of
\bor11
may have their origin in the traditional GCR nucleosynthesis
process.  The key impetus for this re-evaluation
is in new O data in Pop II stars, which suggests that O/Fe 
is not constant with $\feh$.
This implies immediately that Be and B slopes versus Fe 
are not direct indicators of nature of the LiBeB origin,
since O and not Fe should be a direct tracer of this origin.  
At present, the data are too uncertain to {\em require} this scenario,
but it is intriguing to find that this traditional mechanism might be
allowed.

Using recent Pop II data on O/Fe-Fe,
we have analyzed the 
observed BeB-OFe trends.  We find that systematic errors
leave the O data 
too uncertain to allow strong conclusions regarding
the origin of LiBeB.  Specifically, using one (consistent) set
of stellar atmospheric parameters, we find that Be-O shows
a strong hint of a secondary origin, but using
another set of atmospheric parameters, Be-O appears
to have a strong primary component.  Despite these uncertainties, it is
of note that the (better measured) BeB-Fe trends
show a significant sign of a changing B/Be ratio,
and with the recent O/Fe-Fe data can be consistent
with a secondary Be-O relation.

These phenomenological findings are supported by 
detailed numerical calculation of LiBeB galactic chemical
evolution.  We are able to construct both closed
box and outflow models in which \li6BeB arise from
a combination of standard GCR nucleosynthesis and 
the $\nu$-process and can fit the observed trends
(within the systematic errors in the O data and O/Fe-Fe).

We emphasize that just as the {\em present} Pop II data are too
uncertain to decisively test our scenario, they also
cannot unambiguously test
alternative scenarios which posit primary origin for 
Pop II BeB
(e.g., Duncan, Lambert, \& Lemke \cite{dll};
Cass\'{e}, Lehoucq, \& Vangioni-Flam \cite{clv};
Ramaty, Kozlovsky, Lingenfelter, \& Reeves \cite{rklr}).
However, these two basic scenarios differ sharply
in their predictions, and so are exceedingly amenable
to observational tests.  We note that elemental and 
isotopic ratios have very different histories in the
two scenarios, and thus can provide clear signatures.
Specifically, in the scenario we propose
\li{6,7} and \bor11 are all 
primary but Be and \bor10 are secondary.
Consequently, 
key new or improved observations include (in rough order of priority):
\begin{enumerate}

\item Accurate determinations of B/Be within the same stars, 
as a function of metallicity.
The difference in the
observed Be-Fe and B-Fe slopes suggests that this ratio should indeed 
change.
If the evolution of ratio could be well-determined, it would
give the most direct indication (or refutation) of different
primary and secondary origins for Be and B.  

\item Determination of a consistent set of stellar paramters.
One of our main results is that the primary/secondary nature of BeB is
undetermined because of the large systematic uncertainties that arise
from the choice of stellar parameters such as effective temperature,
surface gravity and metallicity.  In particular, if we are to understand
the evolution of these elements with respect to O/H or Fe/H, we need to
have reliable abundance determinations for O/H and Fe/H.  These are not
currently available in the literature.

\item
Measurement of \li6/\be9.  If Be is secondary, then
Pop II \li6/\be9 should be much higher than solar
since Pop II Li is primary;
e.g., in our best models, 
$\li6/\be9 \simeq 30 \simeq 5 (\li6/\be9)_\odot$ at $\feh=-2$.
However, if Be also has primary
component that dominates in Pop II, then \li6/\be9 should 
be constant in Pop II, 
with a change in Pop I as the normal GCR
component becomes important.
Thus, even a rough determination of \li6/\be9 should
constrain the Be origin.

\item Determination of \bor11/\bor10 for Pop II stars.
If the $\nu$-process dominates the B production at early times,
this ratio would provide a strong signature
(e.g., in our best models,
$\bor11/\bor10 \simeq 10$ at $\feh = -2$).
Furthermore, the predicted ratio varies so strongly 
in primary versus secondary models for \bor10
that even rough determinations of this ratio would be worthwhile.
To date, there has only been one attempt at measurement,
we encourage
future attempts to improve the limit.

\end{enumerate}
In deriving trends from any of these or other observations, 
we again stress the importance of using consistent atmospheres.
Since the effect of different model atmosphere inputs can
be large, to compare trends it is essential to
use stellar atmosphere models which are consistent within
and across stars.
 
We conclude by reiterating a main theme of this paper,
that the basic nature of BeB nucleosynthesis
and evolution hangs on Pop II observations of O
in halo stars and a determination of
whether O/Fe is constant or changing in Pop II.
Indeed, the (consistently derived)
correlation of O with Fe, Be, and B is key to
any understanding of light element production in the early Galaxy.  
We strongly 
urge improved observations of O and more 
systematic and consistent study of Pop II abundances
to shed light on this issue.

\acknowledgments
We are very grateful to Elisabeth Vangioni-Flam and 
Michel Cass\'e for helpful discussions and for comments
on an earlier version of this paper. We also thank Evan Skillman for
helpful conversations. BDF acknowledges instructive conversations on boron
isotopes with Luisa Rebull. 
This work was supported in part by
DoE grant DE-FG02-94ER-40823 at the University of Minnesota.

\nobreak

\newpage

\begin{center}
FIGURE CAPTIONS
\end{center}

\begin{enumerate}

\item \label{fig:ofe_fia}
{O/Fe versus $\feh$, plotted for different values of
the SNIa parameter $\fia$.  
The convergence of the curves at $\feh \la -2$ demonstarates
that this regime is dominated by SNII yields.
Models are closed box, and have
an IMF $\propto m^{-2.35}$.
Oxygen abundances are derived from molecular OH lines,
and reflect atmospheric parameters which have been, 
on a star-by-star basis, 
standardized for the LiBeB and O abundances.
{\bf (a)} Data using Balmer line method, 
{\bf (b)} same using IRFM.}

\item \label{fig:BeB-O_cl_b}
Results for closed box model with IMF $\imf \propto m^{-2.65}$;
{\bf (a)} Be versus O and 
{\bf (b)} B versus O.
Pop II abundance data derived using 
the Balmer set of atmosphere parameters (see text).

\item \label{fig:LiBeB-Fe_cl_b}
As in Fig.\ \ref{fig:BeB-O_cl_b}
{\bf (a)} Li and \li6, 
{\bf (b)} Be, and
{\bf (c)} B versus Fe.
Note that we have not included any of the
proposed Pop I sources of \li7 (other than the $\nu$-process);
thus we are unable to fit
the Li behavior at $\feh \ga 0$.
Elemental data are described in the text;
\li6 points combine 
Smith, Lambert, \& Nissen \pcite{sln}, \pcite{sln2};
Hobbs \& Thorburn \pcite{ht}; 
and Cayrel et al.\ \pcite{cay}.  \li6 errors 
reflect uncertainties in the \li6/\li7 ratio only.

\item \label{fig:BBe-Fe_b}
B/Be versus Fe, with 
Pop II abundance data derived using 
the Balmer set of atmosphere parameters (see text).
{\bf (a)} closed box model,
{\bf (b)} outflow model,
{\bf (c)} outflow model shown with IRFM data.

\item \label{fig:BeB-O_E+B-wind_b}
Results for models with 
an enriched wind, and bulk outflow having $\eism = 0.33$.
{\bf (a)} Be versus O and 
{\bf (b)} B versus O.
Pop II abundance data derived using 
the Balmer set of atmosphere parameters (see text).

\item \label{fig:LiBeB-Fe_E+B-wind_b}
As in Fig.\ \ref{fig:BeB-O_E+B-wind_b}
{\bf (a)} Li and \li6, 
{\bf (b)} Be, and
{\bf (c)} B versus Fe.

\item \label{fig:BeB-O_E+B-wind_i}
Results for models with 
an enriched wind, and bulk outflow having $\eism = 0.33$.
{\bf (a)} Be versus O and 
{\bf (b)} B versus O.
Pop II abundance data derived using 
the IRFM set of atmosphere parameters (see text).

\item \label{fig:LiBeB-Fe_E+B-wind_i}
As in Fig.\ \ref{fig:BeB-O_E+B-wind_i}
{\bf (a)} Li and \li6, 
{\bf (b)} Be, and
{\bf (c)} B versus Fe.

\end{enumerate}


 \newpage

\begin{figure}[htb]
\hskip 0.5in
\epsfysize=8truein
\epsfbox{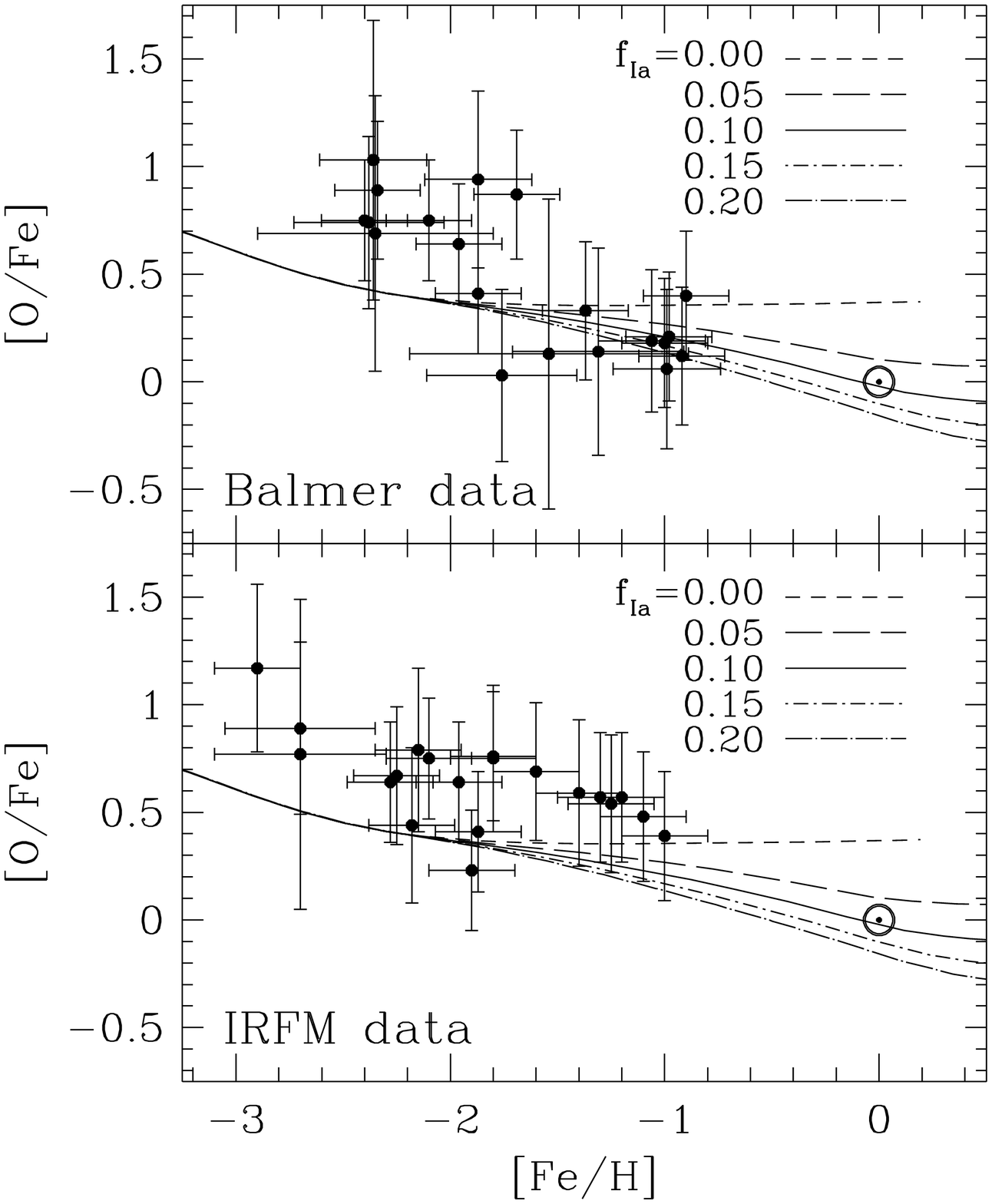}
\end{figure}  
\newpage

\begin{figure}[htb]
\epsfysize=8truein
\epsfbox{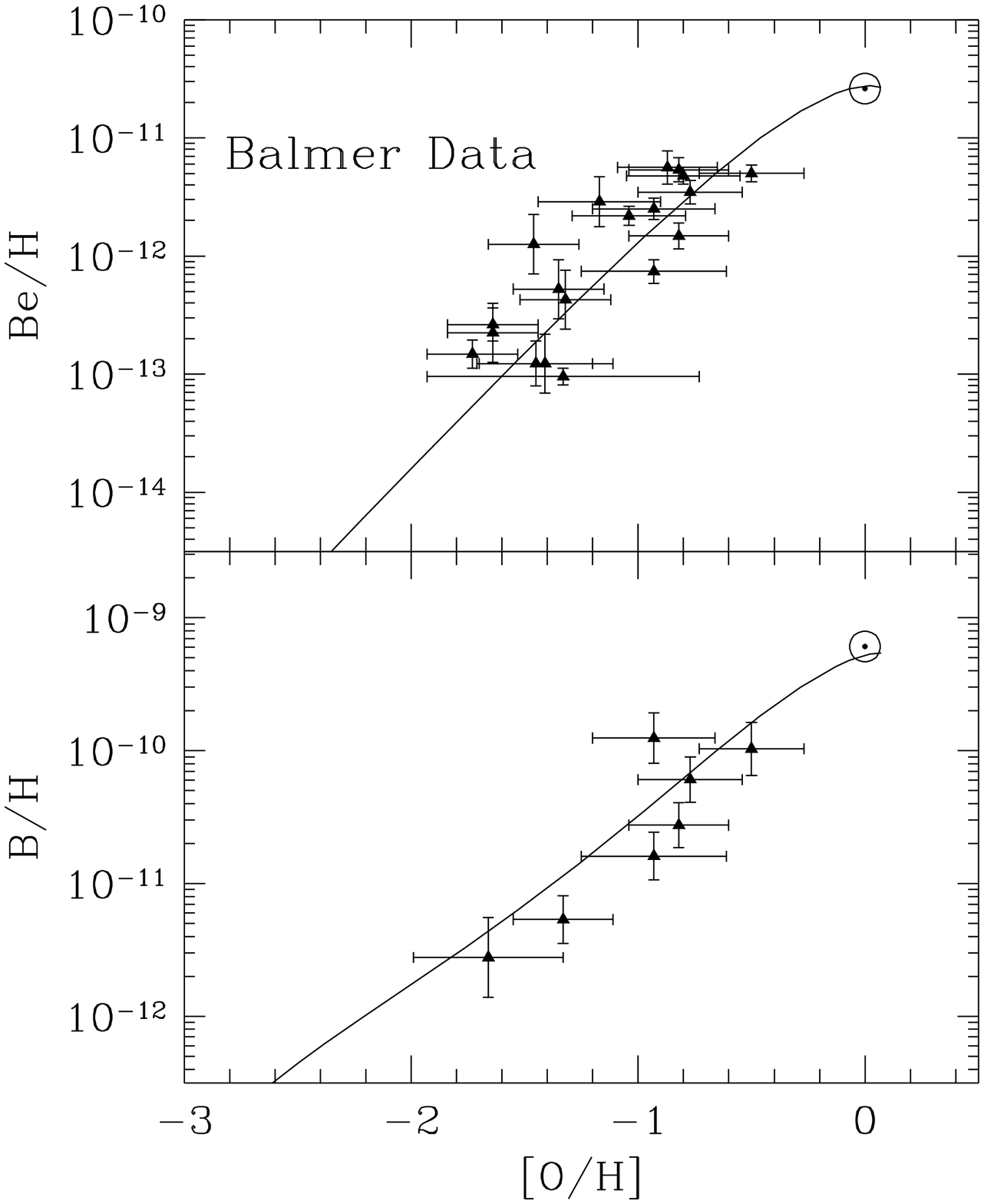}
\end{figure}  

\newpage

\begin{figure}[htb]
\epsfysize=8truein
\epsfbox{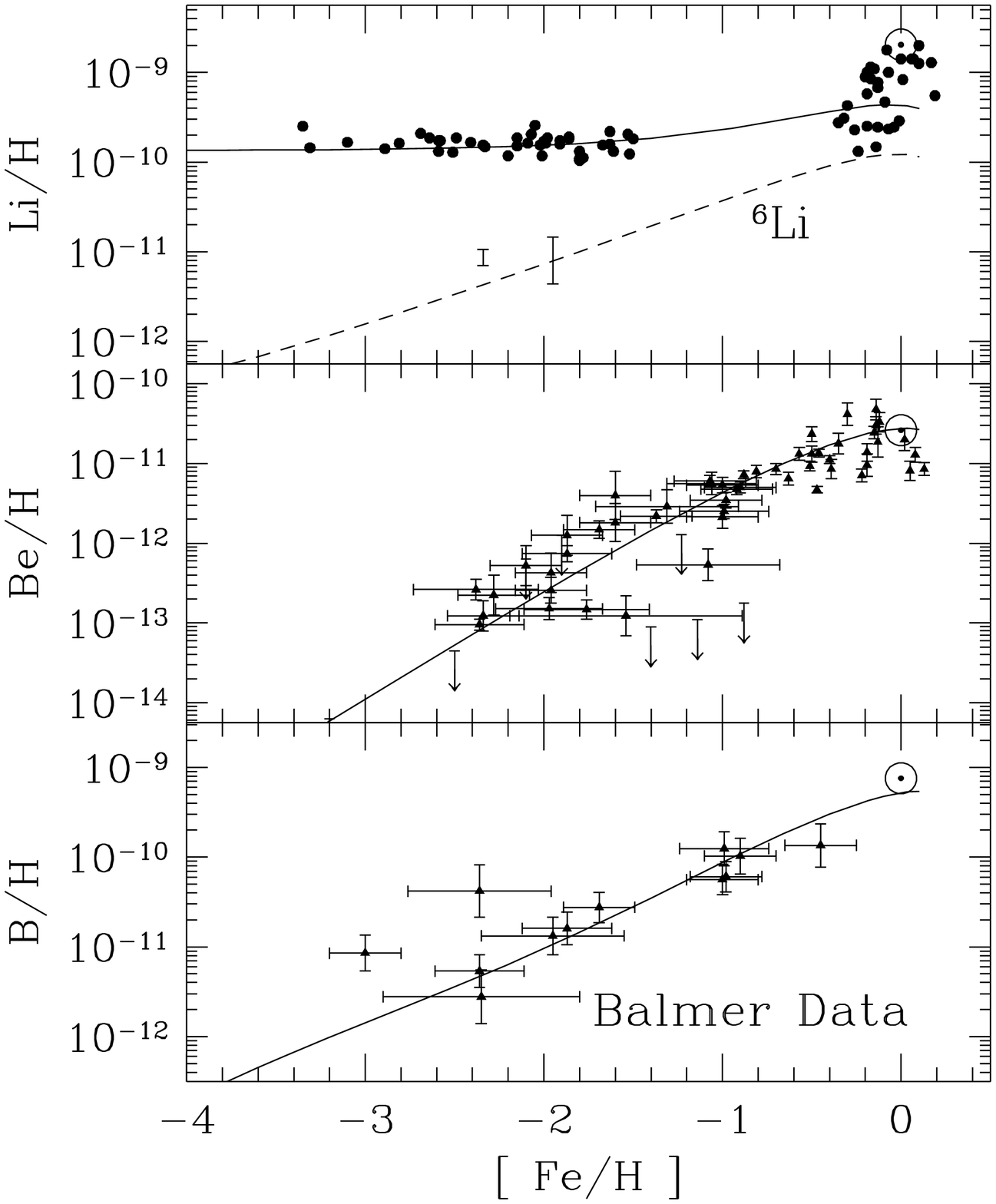}
\end{figure}

\newpage

\begin{figure}[htb]
\epsfysize=8truein
\epsfbox{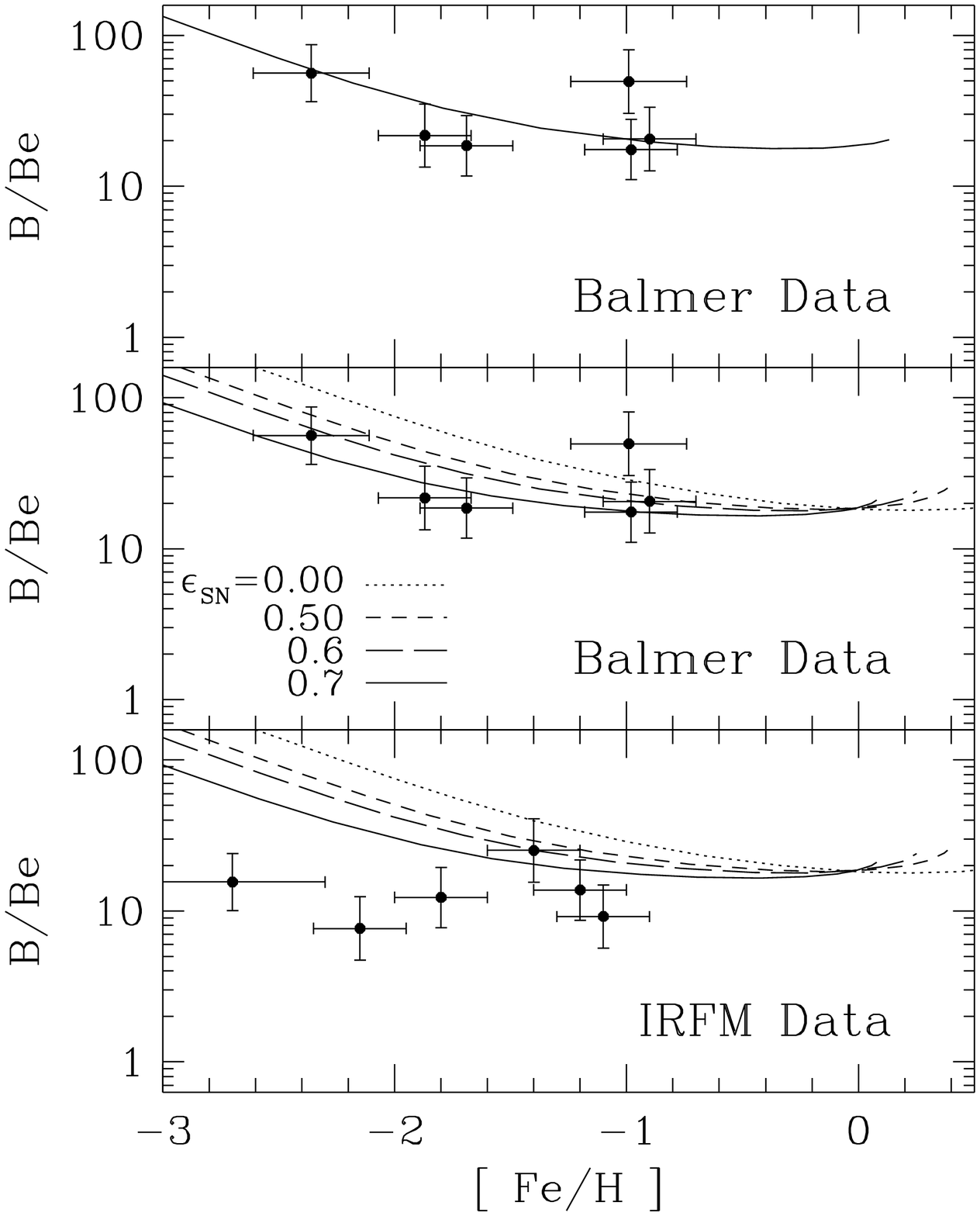}
\end{figure}

\newpage

\begin{figure}[htb]
\epsfysize=8truein
\epsfbox{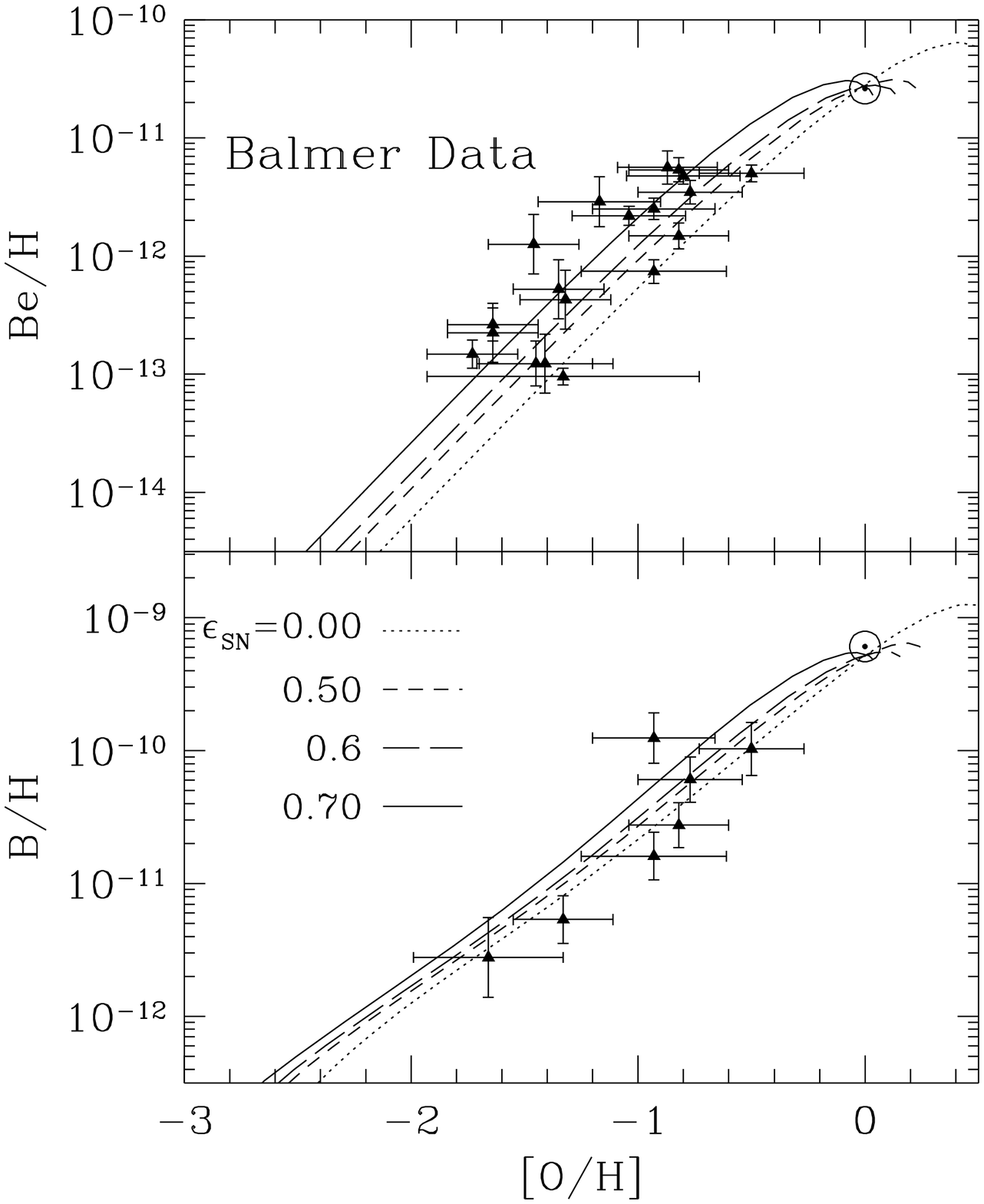}
\end{figure}

\newpage

\begin{figure}[htb]
\epsfysize=8truein
\epsfbox{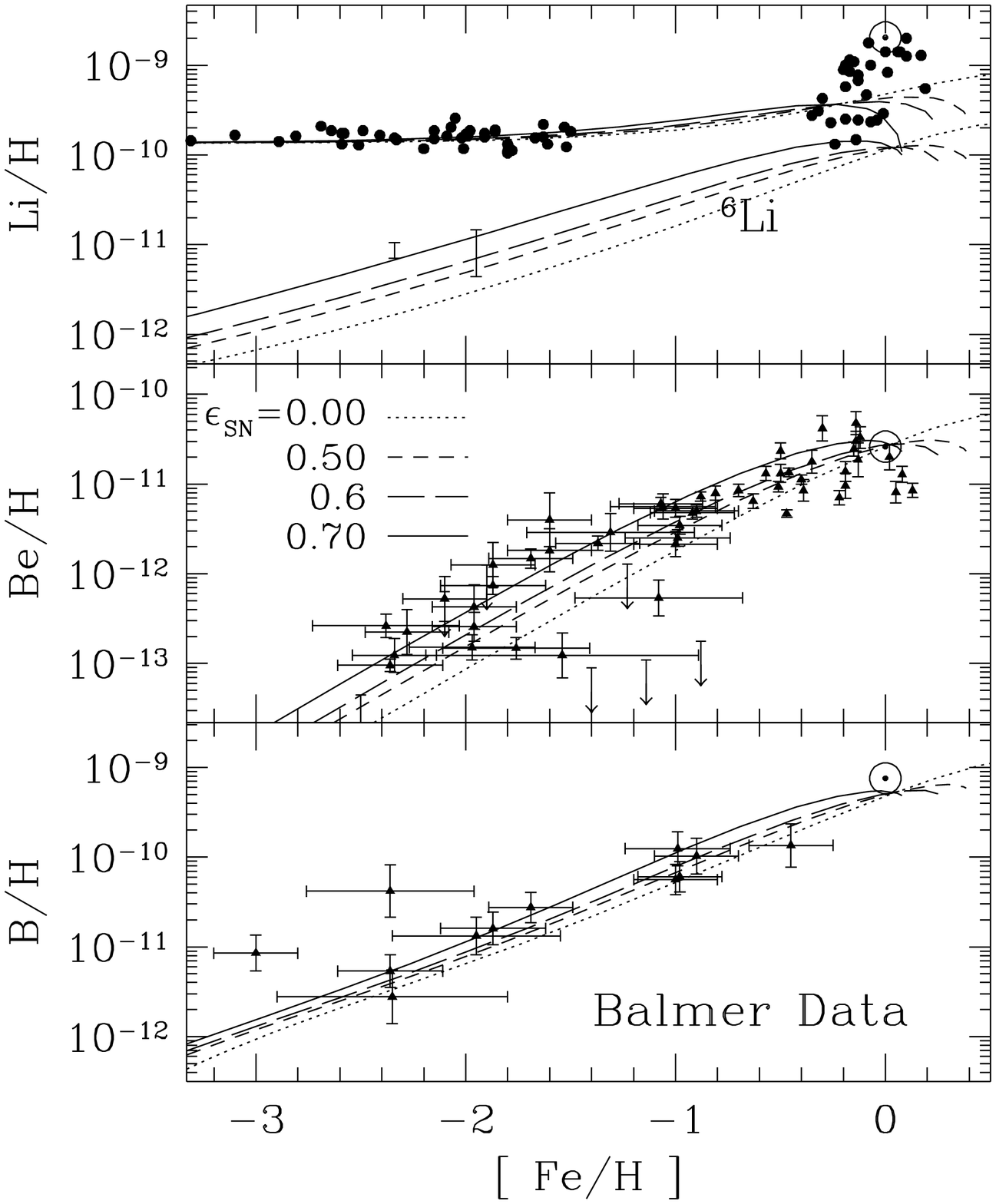}
\end{figure}
\newpage

\begin{figure}[htb]
\epsfysize=8truein
\epsfbox{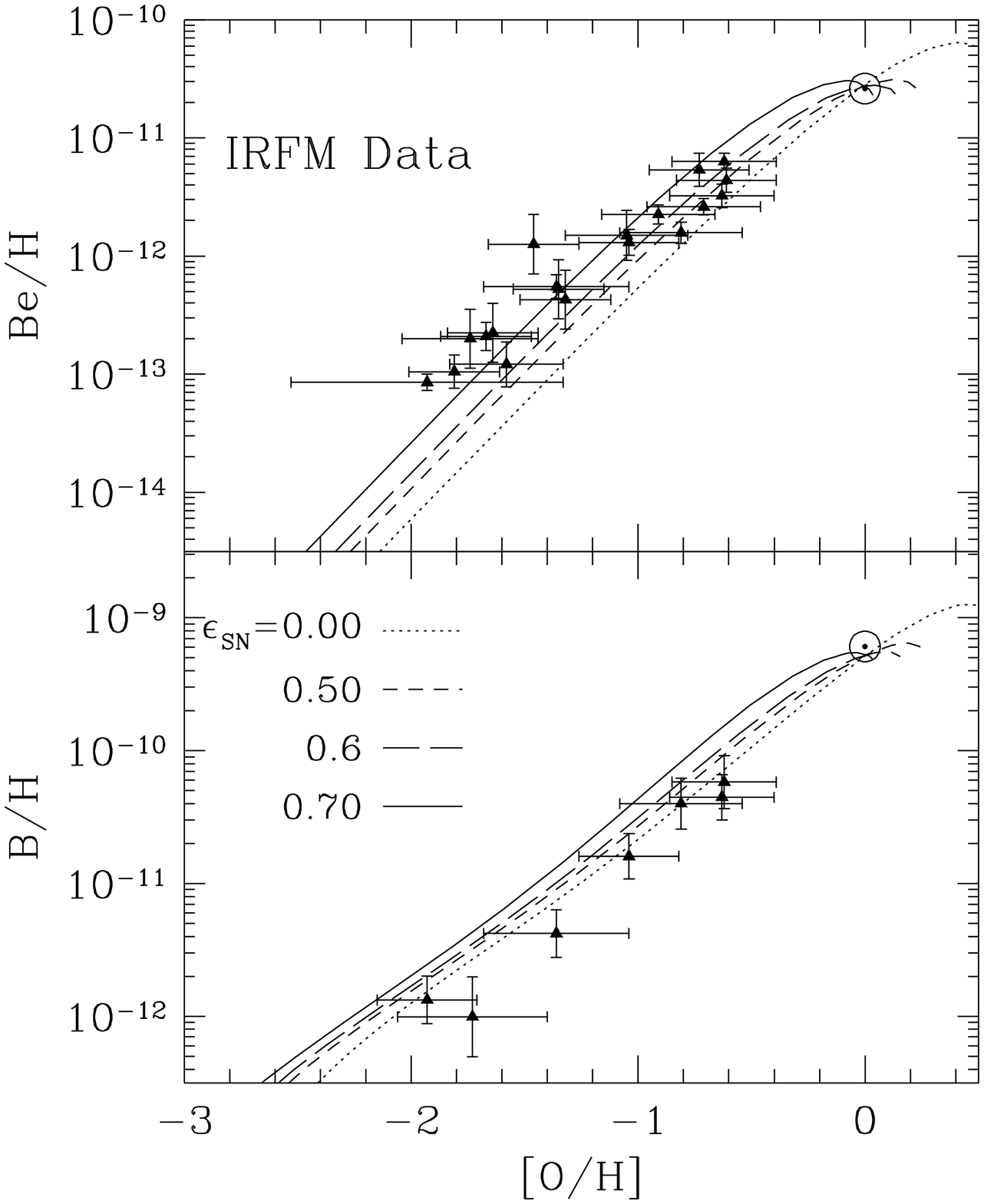}
\end{figure}

\newpage

\begin{figure}[htb]
\epsfysize=8truein
\epsfbox{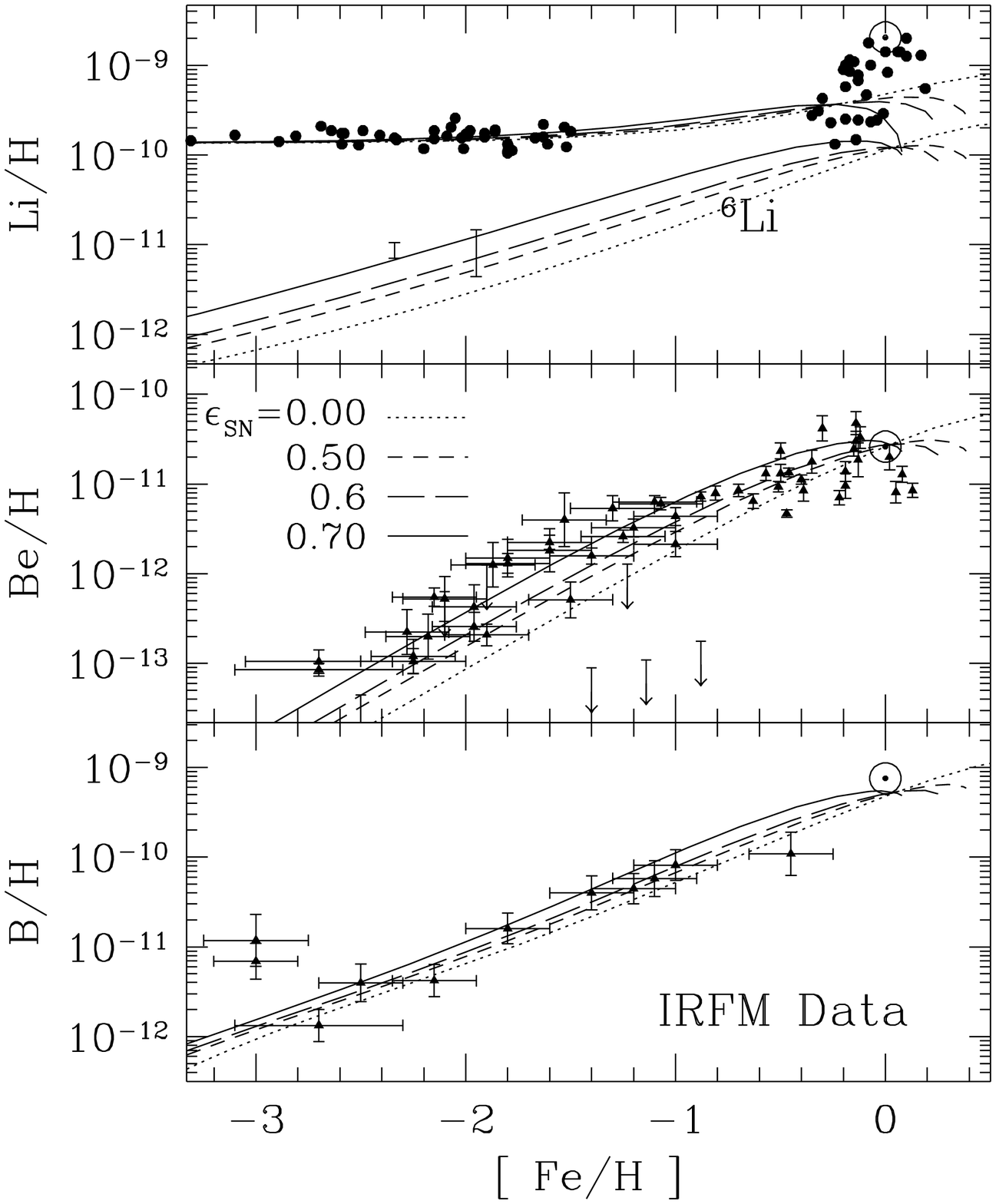}
\end{figure}

\end{document}